\documentstyle[psfig]{mn}

\newif\ifAMStwofonts

\def\kms{\relax \ifmmode {\,\rm km\,s}^{-1}\else \,km\,s$^{-1}$\fi}
\def\ha{\relax \ifmmode {\rm H}\alpha\else H$\alpha$\fi}
\def\hb{\relax \ifmmode {\rm H}\beta\else H$\beta$\fi}
\def\hi{\relax \ifmmode {\rm H\,{\sc i}}\else H\,{\sc i}\fi}
\def\hii{\relax \ifmmode {\rm H\,{\sc ii}}\else H\,{\sc ii}\fi}
\def\h2{\relax \ifmmode {\rm H}_2\else H$_2$\fi}
\def\lha{\relax \ifmmode L_{{\rm H}\alpha}\else $L_{{\rm H}\alpha}$\fi}
\def\shi{\relax \ifmmode \sigma_{{\rm HI}}\else $\sigma_{\rm HI}$\fi}
\def\sh2{\relax \ifmmode \sigma_{{\rm H}_2}\else $\sigma_{{\rm H}_2}$\fi}
\def\degr{\hbox{$^\circ$}}
\def\arcmin{\hbox{$^\prime$}}
\def\arcsec{\hbox{$^{\prime\prime}$}}
\def\deg{\hbox{$^\circ$}}

\def\sec{\hbox{$^{\prime\prime}$}}
\def\fdg{\hbox{$.\!\!^\circ$}}
\def\fs{\hbox{$.\!\!^{\rm s}$}}
\def\farcm{\hbox{$.\mkern-4mu^\prime$}}
\def\farcs{\hbox{$.\!\!^{\prime\prime}$}}
\def\degd#1.#2{ #1\fdg#2 }                 
\def\mind#1.#2{ #1\farcm#2 }               
\def\secd#1.#2{ #1\farcs#2}               
\def\hhh{\ifmmode {\rm ^h}              
         \else {${\rm ^h}$}
         \fi}
\def\sss{\ifmmode {\rm ^s}              
         \else {${\rm ^s}$}
         \fi}
\def\hms#1h#2m#3s{                      
                  \relax
                  \ifmmode #1^{\rm h}\,#2^{\rm m}\,#3^{\rm s}
                  \else \hbox{$#1^{\rm h}\,#2^{\rm m}\,#3^{\rm s}$}
                  \fi
                 }
\def\dms#1d#2m#3s{                      
                  \relax
                  #1\degr\,#2\arcmin\,#3\arcsec 
                 }
\def\hmsd#1h#2m#3.#4s{                  
                      \relax
                      \ifmmode #1^{\rm h}\,#2^{\rm m}\,#3\fs#4
                      \else \hbox{$#1^{\rm h}\,#2^{\rm m}\,#3\fs#4$}
                      \fi
                     }
\def\dmsd#1d#2m#3.#4s{                  
                      \relax
                      #1\degr\,#2\arcmin\,#3\farcs#4
                     }
\def\mag{\relax                          
        \ifmmode ^{\rm m}
        \else $^{\rm m}$
        \fi
       }
\def\magd#1.#2{                          
              \relax
              \ifmmode #1^{\rm m}
                       \hskip-0.55em.\hskip0.22em#2
              \else \hbox{#1$^{\rm m}
                    \hskip-0.55em.\hskip0.22em$#2}
              \fi
             }

\ifoldfss
  \ifCUPmtlplainloaded \else
    \NewTextAlphabet{textbfit} {cmbxti10} {}
    \NewTextAlphabet{textbfss} {cmssbx10} {}
    \NewMathAlphabet{mathbfit} {cmbxti10} {} 
    \NewMathAlphabet{mathbfss} {cmssbx10} {} 
  \fi
  \ifAMStwofonts
    \ifCUPmtlplainloaded \else
      \NewSymbolFont{upmath} {eurm10}
      \NewSymbolFont{AMSa} {msam10}
      \NewMathSymbol{\upi}     {0}{upmath}{19}
      \NewMathSymbol{\umu}     {0}{upmath}{16}
      \NewMathSymbol{\upartial}{0}{upmath}{40}
      \NewMathSymbol{\leqslant}{3}{AMSa}{36}
      \NewMathSymbol{\geqslant}{3}{AMSa}{3E}

    \fi
  \fi
\fi 

\ifnfssone
  \newmathalphabet{\mathit}
  \addtoversion{normal}{\mathit}{cmr}{m}{it}
  \addtoversion{bold}{\mathit}{cmr}{bx}{it}
  \newmathalphabet{\mathbfit} 
  \addtoversion{normal}{\mathbfit}{cmr}{bx}{it}
  \addtoversion{bold}{\mathbfit}{cmr}{bx}{it}
  \newmathalphabet{\mathbfss} 
  \addtoversion{normal}{\mathbfss}{cmss}{bx}{n}
  \addtoversion{bold}{\mathbfss}{cmss}{bx}{n}
  \ifAMStwofonts
    \ifCUPmtlplainloaded \else
      \UseAMStwoboldmath
      \makeatletter
      \new@mathgroup\upmath@group
      \define@mathgroup\mv@normal\upmath@group{eur}{m}{n}
      \define@mathgroup\mv@bold\upmath@group{eur}{b}{n}
      \edef\UPM{\hexnumber\upmath@group}
      \new@mathgroup\amsa@group
      \define@mathgroup\mv@normal\amsa@group{msa}{m}{n}
      \define@mathgroup\mv@bold\amsa@group{msa}{m}{n}
      \edef\AMSa{\hexnumber\amsa@group}
      \makeatother
      \mathchardef\upi="0\UPM19
      \mathchardef\umu="0\UPM16
      \mathchardef\upartial="0\UPM40
      \mathchardef\leqslant="3\AMSa36
      \mathchardef\geqslant="3\AMSa3E
    \fi
  \fi
\fi 

\ifnfsstwo
  \DeclareMathAlphabet{\mathbfit}{OT1}{cmr}{bx}{it}
  \SetMathAlphabet\mathbfit{bold}{OT1}{cmr}{bx}{it}
  \DeclareMathAlphabet{\mathbfss}{OT1}{cmss}{bx}{n}
  \SetMathAlphabet\mathbfss{bold}{OT1}{cmss}{bx}{n}
  \ifAMStwofonts
    \ifCUPmtlplainloaded \else
      \DeclareSymbolFont{UPM}{U}{eur}{m}{n}
      \SetSymbolFont{UPM}{bold}{U}{eur}{b}{n}
      \DeclareSymbolFont{AMSa}{U}{msa}{m}{n}
      \DeclareMathSymbol{\upi}{0}{UPM}{"19}
      \DeclareMathSymbol{\umu}{0}{UPM}{"16}
      \DeclareMathSymbol{\upartial}{0}{UPM}{"40}
      \DeclareMathSymbol{\leqslant}{3}{AMSa}{"36}
      \DeclareMathSymbol{\geqslant}{3}{AMSa}{"3E}
    \fi
  \fi
\fi 

\ifCUPmtlplainloaded \else
  \ifAMStwofonts \else 
    \def\upi{\pi}
    \def\umu{\mu}
    \def\upartial{\partial}
  \fi
\fi

\title[Circumnuclear regions of barred galaxies]{Circumnuclear regions
in barred spiral galaxies\\ I. Near-infrared imaging\thanks{Partly based on
observations made with the NASA/ESA {\it Hubble Space Telescope}, obtained from the data
archive at the Space Telescope Science Institute, which is operated by
the Association of Universities for Research in Astronomy, Inc.  under NASA 
contract NAS 5-26555.}}
\author[D. P\'erez--Ram\'\i rez et al.]{D. P\'erez--Ram\'\i rez$^1$, J. H.
Knapen$^{1,2}$\thanks{Visiting
Astronomer, Canada--France--Hawaii Telescope operated by the National Research
Council of Canada, the Centre National de la Recherche Scientifique de
France and the University of Hawaii.}, 
R. F. Peletier$^{3,4}$, S. Laine$^{1,5}$, R. Doyon$^{6\dagger}$ 
\newauthor and D. Nadeau$^{6\dagger}$
\\
$^1$ Department of Physical Sciences, University of Hertfordshire,
Hatfield, Herts AL10 9AB, UK. E-mail dperez,knapen@star.herts.ac.uk
\\
$^2$ On leave at Isaac Newton Group of Telescopes, Apartado 321, Santa
Cruz de La Palma, E-38700 Spain
\\
$^3$ Department of Physics, University of Durham, South Road, Durham,
DH1 3LE, UK
\\
$^4$ School of Physics and Astronomy, University of Nottingham,
University Park, Nottingham NG7 2RD , UK (present address)
\\
$^5$ Department of Physics \& Astronomy, University of Kentucky,
Lexington, KY 40506-0055, USA (present address)
\\
$^6$Observatoire du Mont M\'egantic and D\'epartement de Physique,
Universit\'e de Montr\'eal, C.P. 6128,\\Succursale Centre Ville,
Montr\'eal (Qu\'ebec), H3C 3J7 Canada, E-mail
nadeau,doyon@astro.umontreal.ca}

\date{Accepted 24 February 2000;
      Received;
      in original form}

\pagerange{\pageref{firstpage}--\pageref{lastpage}}
\pubyear{1999}

\begin{document}

\maketitle

\label{firstpage}

\begin{abstract}

We present sub-arcsecond resolution ground-based near-infrared images of
the central regions of a sample of twelve barred galaxies with
circumnuclear star formation activity, which is organized in ring-like
regions typically one kiloparsec in diameter. We also present {\it
Hubble Space Telescope} near-infrared images of ten of our sample
galaxies, and compare them with our ground-based data. Although our
sample galaxies were selected for the presence of circumnuclear star
formation activity, our broad-band near-infrared images are
heterogeneous, showing a substantial amount of small-scale structure in
some galaxies, and practically none in others.  We argue that, where it
exists, this structure is caused by young stars, which also cause the
characteristic bumps or changes in slope in the radial profiles of
ellipticity, major axis position angle, surface brightness and colour at
the radius of the circumnuclear ring in most of our sample galaxies. In
7 out of 10 {\it HST} images, star formation in the nuclear ring is
clearly visible as a large number of small emitting regions, organised
into spiral arm fragments, which are accompanied by dust lanes.  NIR
colour index maps show much more clearly the location of dust lanes and,
in certain cases, regions of star formation than single broad-band
images.  Circumnuclear spiral structure thus outlined appears to be
common in barred spiral galaxies with circumnuclear star formation.

\end{abstract}


\section{Introduction} 

Observations and modelling of the circumnuclear regions (CNRs) of barred
galaxies give important clues about the nature of active galactic nuclei
(AGN), (circum)nuclear starbursts and the gas dynamics in the central
kpc (e.g. Knapen et al. 1995a,b; Buta \& Combes 1996; Elmegreen et al.
1997; Quillen et al. 1999; Regan \& Mulchaey 1999; see also reviews by
Knapen 1999 and Shlosman 1999).

The presence of non-axisymmetries, such as oval distortions or bars, in
the gravitational potential of a disc galaxy can lead to the infall
of gaseous material, from galactocentric radii of a few kpc, into the
central kpc. The nonaxisymmetry facilitates the removal of angular
momentum from the gas, and directs it into the CNR or possibly the
nucleus (Shlosman et al.  1990; Athanassoula 1992; Phinney 1994). This
gas fuelling process is expected to induce morphological signatures in
the bars and CNRs, which are useful tools in the understanding of the
dynamics and kinematics of galaxies. Dust lanes and nuclear rings appear
to be the dominant circumnuclear features.

\begin{table*}
\hfill
\begin{tabular}{ccrrrcccc}
\hline
Object & Type & Velocity & Distance & Scale & CNR & Nuclear & Date of & Passbands\\
       &      & km s$^{-1}$  &  Mpc   & pc/arcsec & Features & activity & 
observation &\\
\hline 
NGC 1300 & (R')SB(s)bc & 1568 & 20.9 & 101 & Ring&---&05/11/95& $JHK$\\
NGC 1530 & SB(rs)b     & 2461 & 32.8 & 159 & Ring+spiral&---&05/11/95& $JHK$\\ 
NGC 2903 & SAB(rs)bc   &  556 &  7.4 &  36 & Ring&Starburst&05/11/95& $JHK$\\
NGC 3351 & SB(r)b      &  778 & 10.1 &  49 & Ring&Starburst&06/11/95& $K$\\ 
NGC 3504 & (R)SAB(s)ab & 1539 & 20.5 &  99 & Ring+spiral&Starburst&04/02/96 & $JHK$\\
NGC 3516 & (R)SB(s)    & 2649 & 35.3 & 171 & Ring&Seyfert 1.5&06/11/95& $JHK$\\
NGC 3982 & SAB(r)b     & 1109 & 14.8 &  72 & Spiral&Seyfert 2&04/02/96 & $JHK$\\ 
NGC 4303 & SAB(rs)a    & 1566 & 20.9 & 101 & Ring&Seyfert&05/02/96& $JHK$\\ 
NGC 4314 & SB(rs)a     &  963 & 12.8 &  62 & Ring+spiral &--- &03/02/96& $JHK$\\
NGC 4321 & SAB(s)bc    & 1571 & 16.1 &  70 & Ring+spiral&--- &10/06/94& $K$\\
         &             &      &      &     &      &    &04/02/96& $JH$\\
NGC 5248 & SAB(rs)bc   & 1153 & 15.4 &  74 & Ring+spiral&---&04/02/96& $JHK$\\
NGC 6951 & SAB(rs)bc   & 1424 & 19.0 &  92 & Ring+spiral&LINER/Sy&26/09/94& $JHK$\\ 

\hline

\end{tabular}

\caption{Properties of our sample galaxies: morphological classification
(col. 2; data from de Vaucouleurs et al. 1991), velocity (col.3; data
from NED, the NASA/IPAC Extragalactic Database), distance and scale
(col. 4 and 5, from Ferrarese et al. 1996 for NGC~4321 and Graham et al.
1997 for NGC~3351; from the velocity and H$_0$=75 km~s$^{-1}$ Mpc$^{-1}$
for all other objects), the dominant circumnuclear feature as determined
from our data (col. 6), class of nuclear activity (col. 7; data from
NED), date of run and passbands (col. 8 and 9). All images were obtained
with MONICA on the {\it CFHT} except the $K$-image of M100, which was obtained
at UKIRT (Knapen et al. 1995a).}

\end{table*}

Dust lanes have been interpreted as the location of shocks in the gas
flow. Thus, the morphology of the dust distribution can reveal
characteristics of the dynamics of these galaxies (Athanassoula 1992)
and of the kinematics of the gas. According to Athanassoula's models,
the degree of curvature of the dust lanes is a direct indicator of the
strength of the bar for bars which are not very strong. Nuclear rings,
which are usually sites of active star formation (SF), occur at the
location of strong density enhancements in the gas, where the bar-driven
inflow of gas slows down in the vicinity of inner Lindblad resonances
(ILRs; Athanassoula 1992; Heller \& Shlosman 1994; Buta \& Combes 1996;
Shlosman 1999). SF may result from the gas becoming gravitationally
unstable (Elmegreen 1994, 1997), or from triggering in miniature spiral
arms (Knapen et al. 1995b, 2000).

Whereas imaging in the blue, ultraviolet, or in spectral lines like
H$\alpha$ traces the massive SF (e.g. Sersic \& Pastoriza 1967; Pogge
1989a,b; Knapen et al. 1995a; Maoz et al.  1996; Colina et al. 1997),
near-infrared (NIR) imaging offers very important advantages. Firstly,
the presence of substantial and mostly unquantified amounts of
extinguishing (absorbing and scattering) dust in the CNRs hampers the
interpretation of imaging at especially the UV and optical
wavelengths. NIR, particularly $K$-band, emission is much less
susceptible to extinction by dust (a factor of 10 when comparing $K$ to
$V$), whereas colour index images such as $I-K$ (Knapen et al. 1995a) or
$J-H$ are clear morphological dust indicators due to the relatively
modest changes in those colours caused by different stellar populations.
Secondly, NIR observations are of particular importance to the study of
the CNRs because they allow a better determination of the mass
distribution (excluding dark matter). NIR imaging is more sensitive to
light primarily from cool giants and dwarfs which dominate the mass or
are at least directly proportional to it. However, as shown in the
literature and by us in this paper, there can be significant, or even
dominant, contributions from young stars to the NIR emission even in the
$K$-band, so for objects with strong SF any estimate of the mass
distribution made from NIR maps must take the varying $M/L$ ratio into
account (Knapen et al. 1995a,b; Elmegreen et al. 1997; Wada, Sakamoto \&
Minezaki 1998; Ryder \& Knapen 1999).

The atlas of images presented in this paper (Paper I) represents the
first results of a programme which studies the circumnuclear structures
that appear at small scales (a few hundred parsecs to a kpc) and their
connection with the global disc structure in a sample of 12 barred
spiral galaxies. We present $J, H$ and $K$-images at subarcsec
resolution for all 12 sample galaxies, and {\it Hubble Space Telescope}
({\it HST}) archive $H$-band images for 10 of them. In Paper II
(P\'erez--Ram\'\i rez \& Knapen 2000a), we present an accompanying
optical data set of broad-band and H$\alpha$ images of the complete
discs of all our sample galaxies, while in Paper III (P\'erez--Ram\'\i
rez \& Knapen 2000b) the morphological information is interpreted in
terms of the structure and evolution of CNRs of barred galaxies.

In Section 2, we describe the observations and data reduction
techniques. The imaging data are presented as a series of multi-panel
figures, which show broad-band images, colour index images highlighting
dust and SF features against the dominant stellar luminosity, and radial
profiles of ellipticity, position angle, surface brightness and
colour. The results are summarized in Section 3 for individual galaxies,
and briefly discussed in Section 4, while concluding remarks are given
in Section 5.

\setcounter{figure}{0}

\begin{figure*}
\vspace{1cm} 
\caption{{\it HST} NICMOS images of the central areas of our sample
galaxies obtained using the F160W filter (comparable to $H$-band). The
regions shown are 19\sec\ squared in all cases. The grey scales are
stretched in such a way as to emphasize the morphology of the CNR. Only
NGC~3516 is shown with a logarithmic grey scale stretch, all others linear.
Black and white border regions are due to rotation of the images and the
orientation of the object on the array. The orientation of the images is
the same as of those in Fig.~2, i.e., N is up, E to the left.}
\end{figure*}
 
\section{Sample Selection, Observations and Data Reduction}

\subsection{The sample}

\setcounter{figure}{1}

\begin{figure*}
\psfig{figure=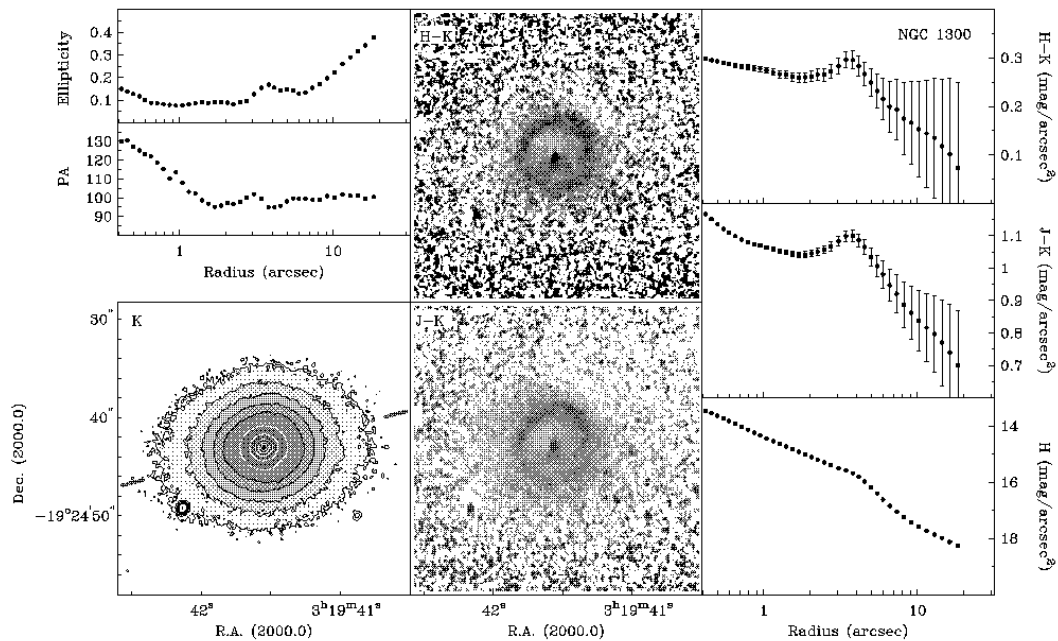,height=12cm,angle=270}
\caption{Images and photometric profiles for the individual galaxies.
On the upper left ellipticity $(1-b/a)$ and major axis position angle
(in degrees) are plotted as a function of major axis radius. The
position angle has been measured from N through E.  The lower left panel
is a contour plus grey-scale plot of the $K$-band image of the
galaxy. Contours start at $K$=17.5\,mag\,arcsec$^{-2}$, with intervals
of 0.5 mag\,arcsec$^{-2}$.  The coordinates are J2000.0, obtained from
the NASA/IPAC Extragalactic Database (NED). The orientation of the major
axis of the main bar, as determined from the optical imaging in
Paper~II, is indicated whenever appropriate by a pair of tickmarks near
the border of the lower left panel. The middle two panels are grey-scale
representations of $J-K$ and $H-K$ colour index images with a grey-scale
range of 0.5 mag\,arcsec$^{-2}$.  Redder colours are indicated by darker
shades. The right panels show the $H$-band surface brightness profile
(bottom), and the $J-K$ (middle) and $H-K$ (top) colour profiles. The
colours and the $H$-band surface brightness have been photometrically
calibrated using aperture photometry in the literature (see Section
3). {\bf a.} NGC 1300}
\end{figure*}

The main selection criterion for our sample is the presence of a bar and
evidence of some circumnuclear structure associated with it, such as
rings, nuclear bars or regions of SF. Furthermore, the
sample galaxies should be nearby, bright and observable from the
northern hemisphere. Some of the sample galaxies are taken from the
lists published by Sersic \& Pastoriza (1967) and Pogge (1989a,b).

The sample can be considered more anecdotal than complete in any
sense. However, it does significantly increase the sample size when
compared to published NIR studies of CNRs in barred galaxies, which
rarely include more than three objects and usually discuss only one
(e.g. Knapen et al. 1995a,b; Elmegreen et al. 1997; Laine et al. 1999;
Ryder \& Knapen 1999). With our new imaging, we also improve the
spatial resolution, and show different colour index maps, $J-K$
(primarily outlining dust extinction) and $H-K$ (additionally
indicating the possible emission due to hot dust in the case of the
more active sample galaxies). Regan \& Mulchaey (1999) show {\it HST}
1.6$\mu$m and optical-infrared (basically $R-H$) colour index images
of a sample of Seyfert (Sy) galaxies, including a number of CNR
galaxies. Two of the galaxies in their sample are also included in our
sample (NGC 3516 and NGC 3982). The $R-H$ colour, however, is much
more sensitive to changes in stellar populations than the NIR colours
used in our paper. For 10 of our sample galaxies, we
have retrieved, in addition to our own data, $H$-band images from the
{\it HST} archive (Fig.~1), and we comment on the high-resolution NIR
morphology of those objects. In Table 1 we give some information about
the classification of the galaxies of our sample, the predominant
circumnuclear feature, as determined from the data, and the type of
nuclear activity, if known.

\setcounter{figure}{1}

\begin{figure*}
\psfig{figure=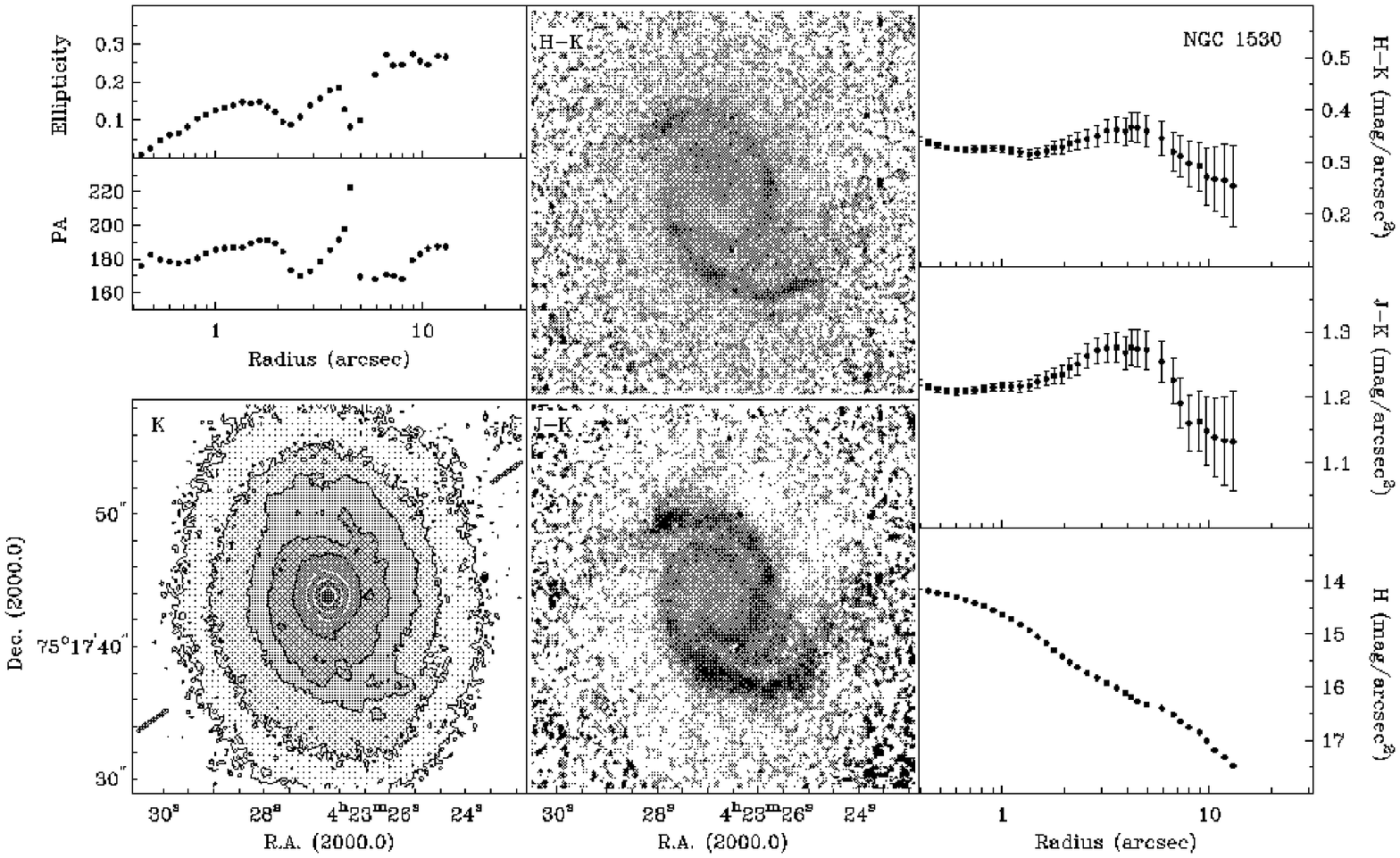,height=12cm,angle=270}
\caption{{\bf b.} NGC 1530}
\end{figure*}

\subsection{Ground-based observations}

We have obtained NIR broad-band images in the $J$ (1.25$\mu$m), $H$
(1.65 $\mu$m) and $K$ (2.2 $\mu$m) bands of the 12 barred galaxies
that make up our sample, during three observing runs: 1994 September
26, 1995 November 5--6, and 1996 February 3--5. The observations were
made with the 3.6 m Canada--France--Hawaii Telescope ({\it CFHT}). We used
the Montreal NIR camera (MONICA; Nadeau et al. 1994), equipped with a
256 x 256 pixel HgCdTe array detector with a projected pixel size of
\secd 0.{248}.

Since the sky background in the NIR is bright and changes rapidly, it
is essential to obtain sky frames frequently. We observed the sky
usually just before and after each galaxy observation.  Frames
typically consisted of four co-added individual exposures.  Separate
frames with the nucleus of the galaxy at slightly offset positions
were co-added in the reduction process to produce the final mosaic
images.

The weather was clear and generally photometric thoughout these
nights, and the seeing values as measured from our final images were
\secd 0.7 -- \secd 1.0.

\setcounter{figure}{1}

\begin{figure*}
\psfig{figure=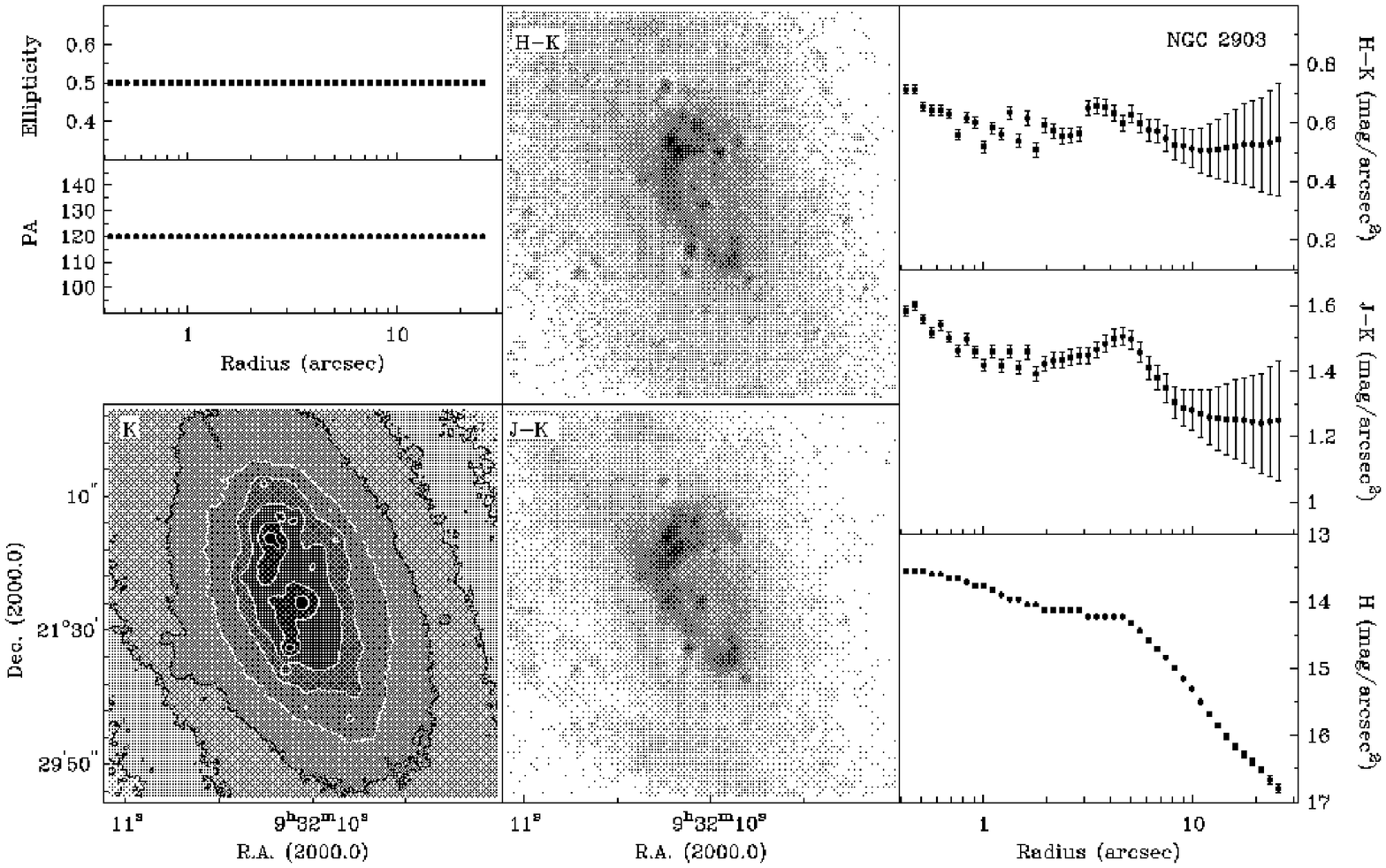,height=12cm,angle=270}
\caption{{\bf c.} NGC 2903, ellipticity and position angle were fixed 
in the fitting process}
\end{figure*}

\subsection{Data reduction} 

The main steps in the reduction of the NIR data include subtracting
the sky background, interpolating across known bad pixels, and
registering and combining the images. The data reduction was done
partly with private programmes and partly with standard {\sc iraf}
routines.

We first combined four sky exposures taken at different positions on the 
sky immediately before and
after a series of galaxy exposures by median-averaging them, while
iteratively rejecting those values in the averaged sky exposure that
deviate by more than 3$\sigma$ from the average value in that frame.
This procedure allows the automated rejection of any star images that
might have been present in the sky exposures, which were observed by
offsetting the telescope blindly to a position a few arcmin from the
galaxy centre. We then subtracted the averaged sky image from the
relevant galaxy images.

A close inspection of the sky-subtracted galaxy images of the 1996 run
showed the presence of periodic, horizontal lines which could be traced
back to electronic crosstalk in the system. This raises the noise level
in these images. In order to correct the problem, we used a Fourier
transform technique to locate the frequencies of the maximum intensity
in the images, and filtered them out. This resulted in a considerable
improvement in most images, although in a few images interference
stripes are still notable at low levels. 

After flat-fielding with dome flats, and masking the unreliable pixels
in the array, we combined the sky-subtracted and, where appropriate,
de-striped images.  Ideally, one would like to locate one or more common
field stars in the images to achieve the most accurate alignment
possible. Unfortunately, the field of view of our images is small and in
most cases we could only use the nucleus of the galaxy itself to
determine the reference position.  The individual images were thus
registered and shifted to a common position (in all cases accurate to a
fraction of the pixel size) and averaged. In the process, the pixel size
was halved to increase sampling (reduced from 0$\farcs 248$ to 0$\farcs
124$). Finally, we rotated the images by an angle of 88.8$\deg$ in order
to obtain the correct north (up), east (left) orientation.

\setcounter{figure}{1}

\begin{figure}
\psfig{figure=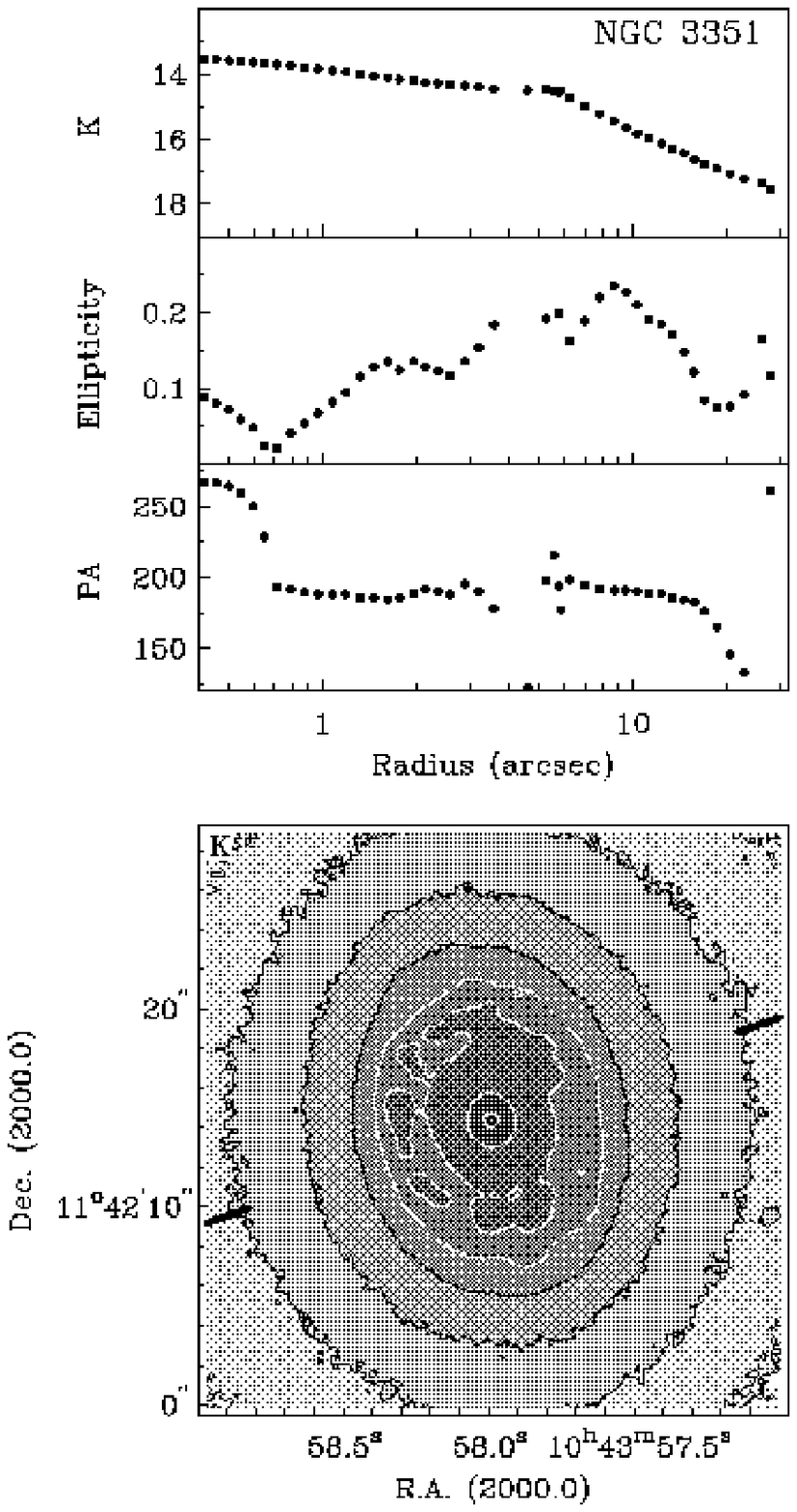,width=8.5cm}
\caption{{\bf d.} NGC 3351}
\end{figure}

\subsection{Photometric Calibration}

Although we subtracted the background sky as described in Section 2.3,
we cannot check on how accurate our sky subtraction is since the sky is
variable, and the frame is small compared with the galaxy.  This makes
it very hard to estimate to what level the flux registered in the
photometric aperture is influenced by any residual background emission.
For that reason, we have used aperture photometry from the
literature. By calibrating with multiple aperture measurements we can
better estimate the value of the sky and the efficiency of the
system. If only one useful aperture is available, we can use only the
relatively bright part of the image. We were able to find the aperture
photometry in $J$, $H$ and $K$-bands for all but one of the galaxies,
NGC 1530. For this galaxy, we used the average magnitude offset as
determined for the rest of the galaxies. Literature photometry sources
include Glass (1976), Aaronson (1977), McAlary, McLaren \& Crabtree
(1979), Balzano \& Weedman (1981), Willner et al. (1985) and Spinoglio
et al. (1995).

To check the quality of our photometry, we also made comparisons with
data from the literature. We selected values from Glass (1984), Cidziel,
Wynn-Williams \& Becklin (1985), Devereux, Becklin \& Scoville (1987)
and Hunt et al. (1994).  The average absolute difference between our
results and the results from different authors was 0.12 mag.

Direct comparison of the surface brightness profile in the $H$-band
and the $H-K$ colour profile plots was possible for two of our sample
galaxies, NGC 3516 and NGC 3982, which are in common with the Peletier
et al. (1999) sample.  The photometric calibration procedures
used were slightly different, but the agreement in both cases was good.

\setcounter{figure}{1}

\begin{figure*}
\psfig{figure=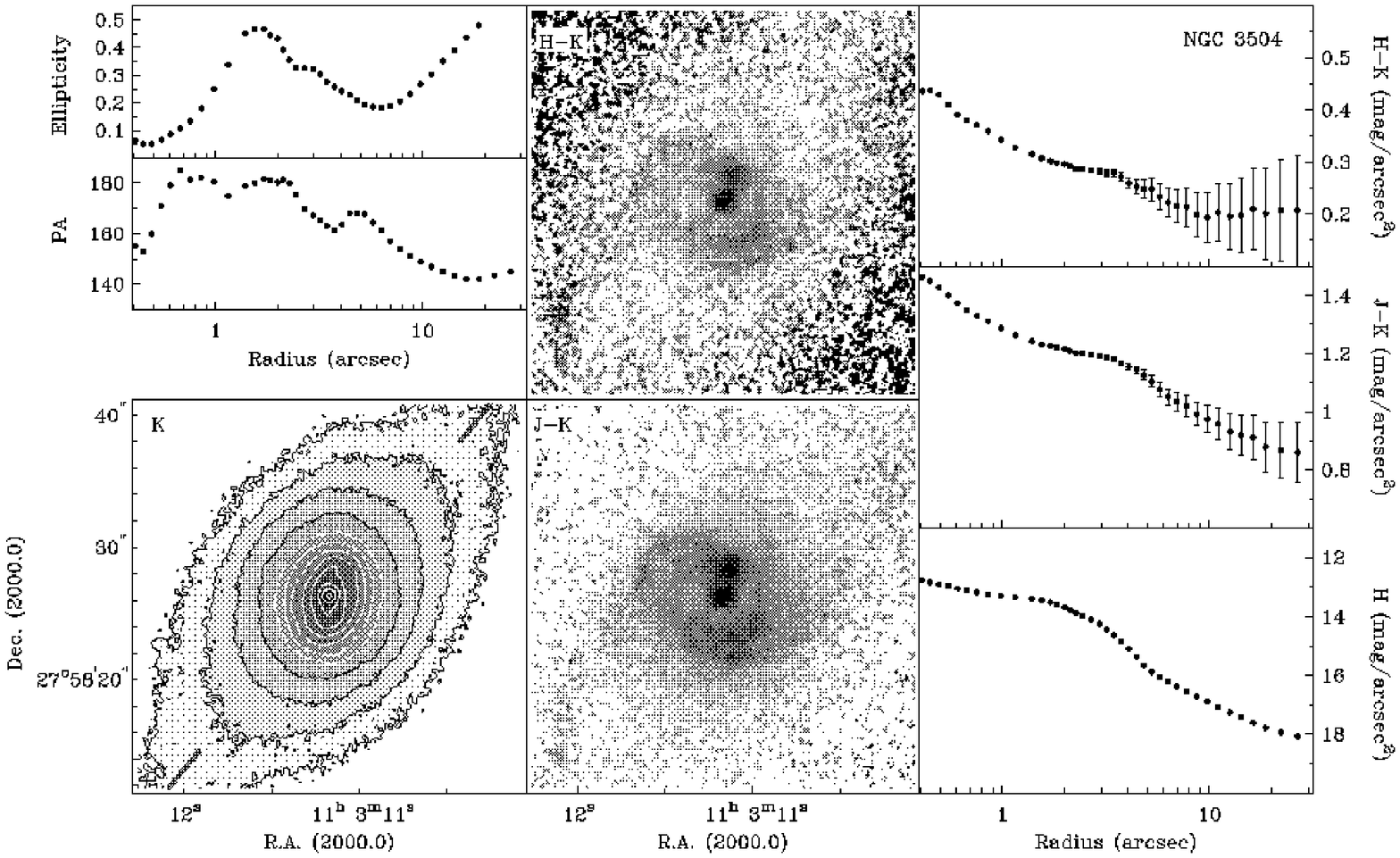,height=12cm,angle=270}
\caption{{\bf e.} NGC 3504}
\end{figure*}

\subsection{Colour index maps and radial profiles}

Colour index maps were created after assuring that the relevant pairs of
images were at the same pixel scale and orientation and at comparable
resolution before combining them.

After calibrating the images, we used {\sc galphot} (J{\o}rgensen, Franx
\& Kjaergaard 1992) to fit ellipses to the images. The centre position,
ellipticity and the position angle of the fitted ellipses were allowed
to change freely as a function of radius for the $H$ passband. The
values for these parameters were then used as input for the fits to the
images in other passbands, in order to ensure that we produce reliable
colour profiles. In the case of NGC~2903, the position angle and
ellipticity were kept constant because the large amount of structure
present in the core did not allow a meaningful fit to these
parameters. We thus produced plots (shown in Fig.~2) of the surface
brightness of the $H$-band as a function of radius, and colour profiles
in $J-K$ and $H-K$ for all the galaxies which were imaged in these
bands. Errorbars in these plots indicate the uncertainty in the
determination of the sky background. We have estimated these
uncertainties using the procedure described by Peletier et al. (1999),
where 1$\sigma$ errorbars correspond to $\mu_{H}$= 21.5 mag
arcsec$^{-2}$, $\mu_{J}$= 21.2 mag arcsec$^{-2}$ and $\mu_{K}$= 20.5 mag
arcsec$^{-2}$ (about 0.1\% of the sky background in $H$ and $K$ and 0.3
\% in $J$).

In Fig.~2, we show for each galaxy (except NGC~3351 for which we only
obtained a $K$-band image) greyscale representations of the $K$
broad-band and $J-K$ and $H-K$ colour index images, radial profiles of
position angle and ellipticity as determined from the fit to the
$H$-band image ($K$ for NGC~3351), and radial surface brightness ($H$)
and colour ($J-K$ and $H-K$) profiles. We tabulated the values plotted
as a function of radius, but publish the tables in electronic form
only\footnote{The data tables are available electronically from the
Centre de Donn\'ees astronomiques de Strasbourg (CDS), on:
ftp://cdsarc.u-strasbg.fr/pub/cats/J/MNRAS/volume/first\_page .}. They
are also available from the authors.

\setcounter{figure}{1}

\begin{figure*}
\psfig{figure=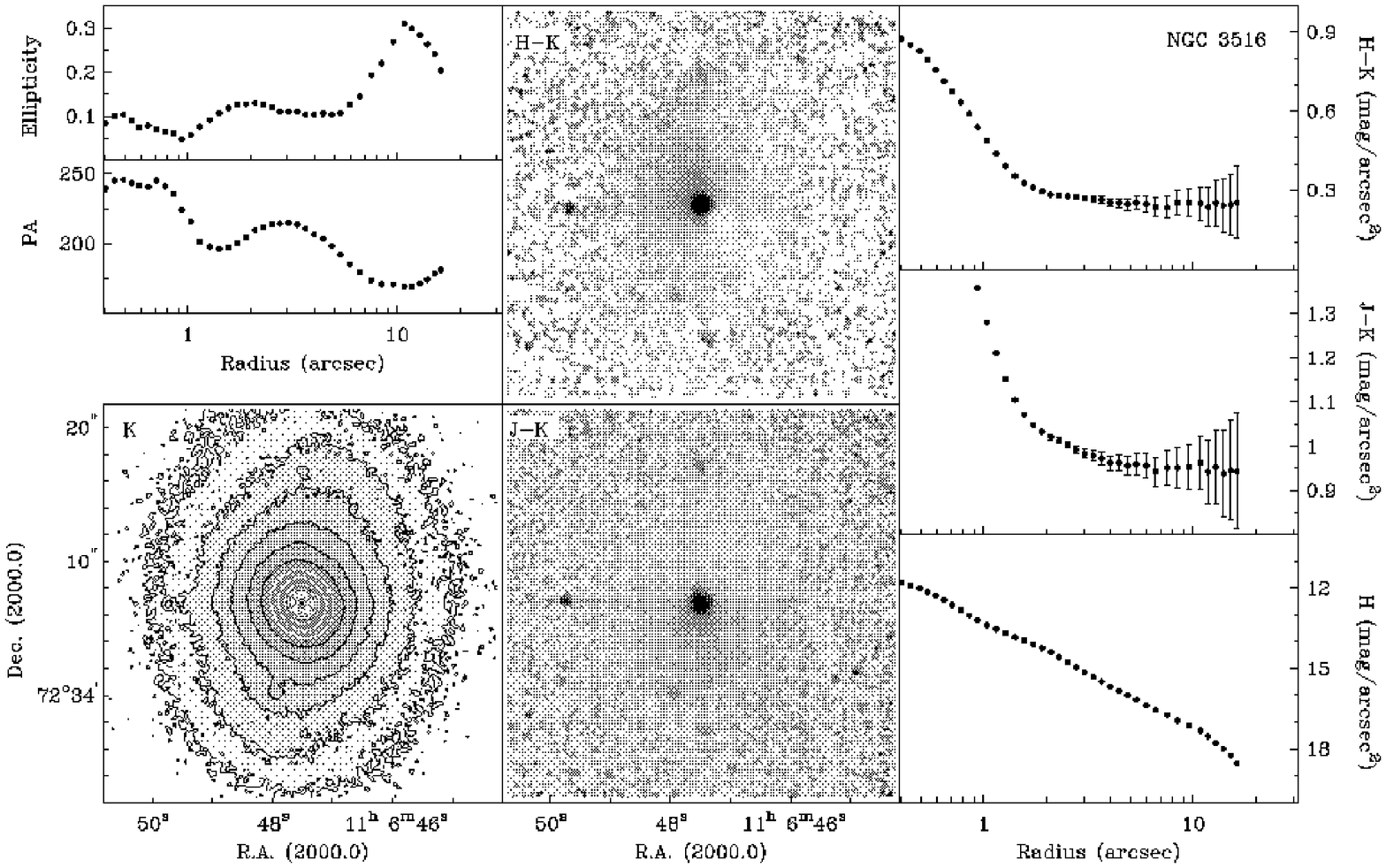,height=12cm,angle=270}
\caption{{\bf f.} NGC 3516}
\end{figure*} 

\subsection{{\it HST} NIR imaging}

For 10 of our 12 sample galaxies, NICMOS imaging with the {\it HST} is
available from the {\it HST} archive. We retrieved the re-reduced F160W
(comparable to $H$-band) images from the archive. However, we improved
the quality of some of these images by doing additional data reduction
to remove artifacts.  We relied on header information to place the
images on an astrometrically correct grid. The images are all taken with
the NIC2 camera, with a pixel size of \secd 0.075. In most cases, the
image retrieved from the archive is a combination of several individual
exposures. For NGC~3516 and NGC~3982, however, one single exposure was
available, and these images in fact improved most due to our additional
data reduction. Regan \& Mulchaey (1999) published two of these images
(NGC 3516 and NGC 3982). We show the central areas of all images in
Fig.~1, with the same scale and orientation. In most of the objects, a
wealth of detail can be seen in the CNR. The emission usually coincides
with the location of the SF ring, and most of the individual bright
knots are due to regions of current SF. Dust lanes and/or SF regions
often outline spiral-like patterns, which will be discussed in more
detail below, and in Paper~III.

\setcounter{figure}{1}

\begin{figure*}
\psfig{figure=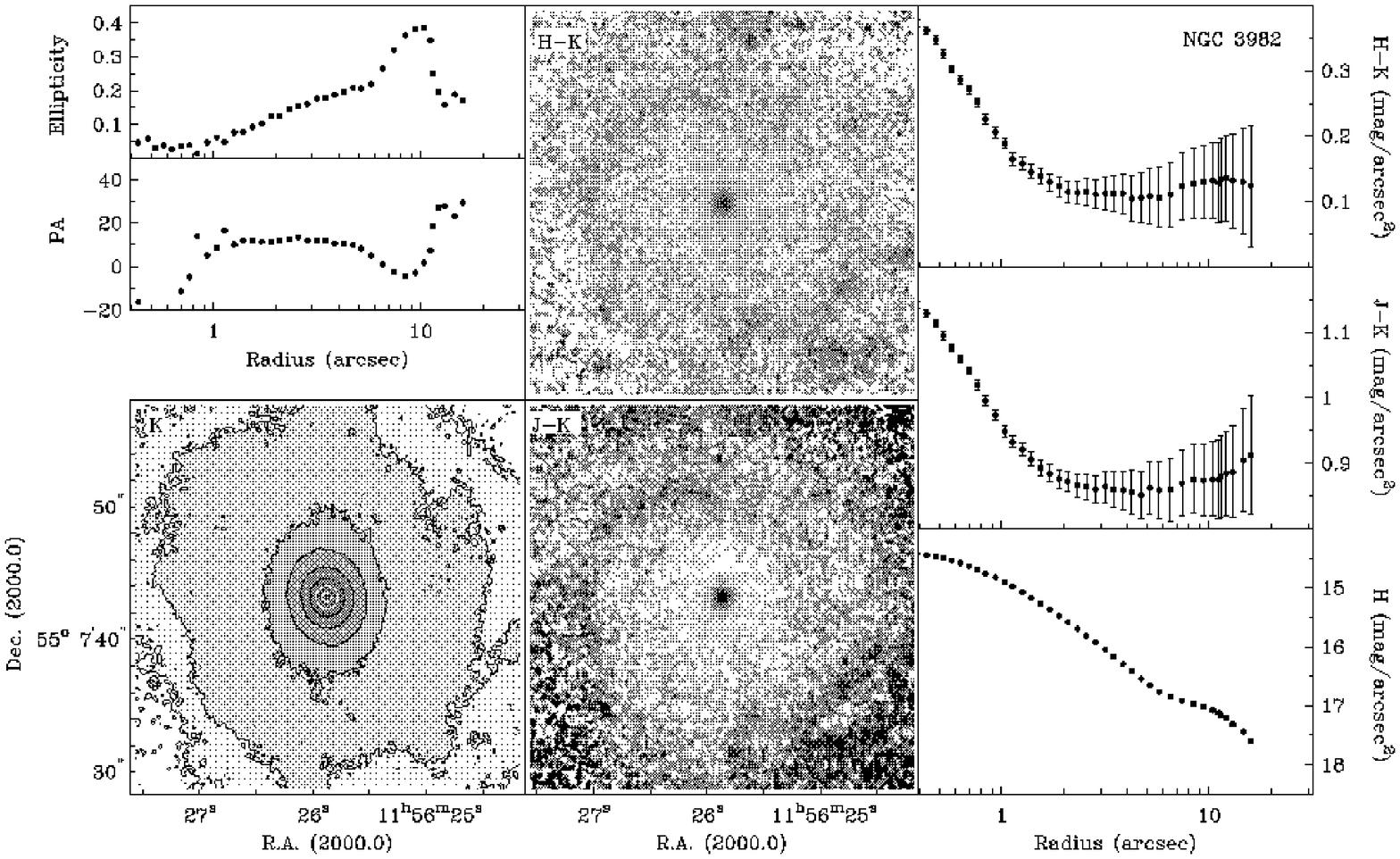,height=12cm,angle=270}
\caption{{\bf g.} NGC 3982}
\end{figure*} 

\section{Results on individual galaxies}

In this section, we describe some of the results of the NIR imaging of
our sample galaxies, as shown in detail in Fig.~1 ({\it HST} NIR images)
and Fig.~2 (ground based multi-band images and profile fits).  A more
systematic study of parameters derived from these data in combination
with optical imaging of the complete galaxy discs is forthcoming (Papers
II and III).

\subsection{NGC 1300}

Our broad-band NIR images are remarkably smooth, and do not show any
structure in the CNR (Fig.~2a).  The colour index images, however, show
a red ring-like structure, possibly outlining a single spiral arm that
departs from the nucleus towards the west side of the image and
continues to wrap around the nucleus until it closes in a ring. This
ring appears more continuous and broader in the northern part than in
the south where it looks patchy. The feature seen in our NIR colour
index map cannot be an artifact of combining two images with slightly
different spatial resolution, because such an artifact would show up as
a complete ring, whereas the observed red feature is not at constant
galactocentric radius. Pogge (1989a) saw an incomplete nuclear ring in
\ha\ emission with a number of distinct ``hot spots''.

The {\it HST} $H$-band image resolves the nuclear ring into a series of
tightly wound spiral armlets, outlined in dust and stellar
populations. The ring region is of a relatively low amplitude compared
to the central bulge component, in contrast to nuclear rings in other
objects, e.g. NGC~3351, as can be observed in Fig.~1.

The location of the ring corresponds to a bump in all radial
profiles (i.e., surface brightness, colour, ellipticity and position
angle). The $J-K$ colour of the ring is redder by 0.1 magnitudes than
the background. No isophotal twists are seen in the ellipticity and
position angle profiles or in the contour map of the $K$-band
image. Thus we confirm that there is no evidence for a nuclear bar
(Regan \& Elmegreen 1997).

\setcounter{figure}{1}

\begin{figure*}
\psfig{figure=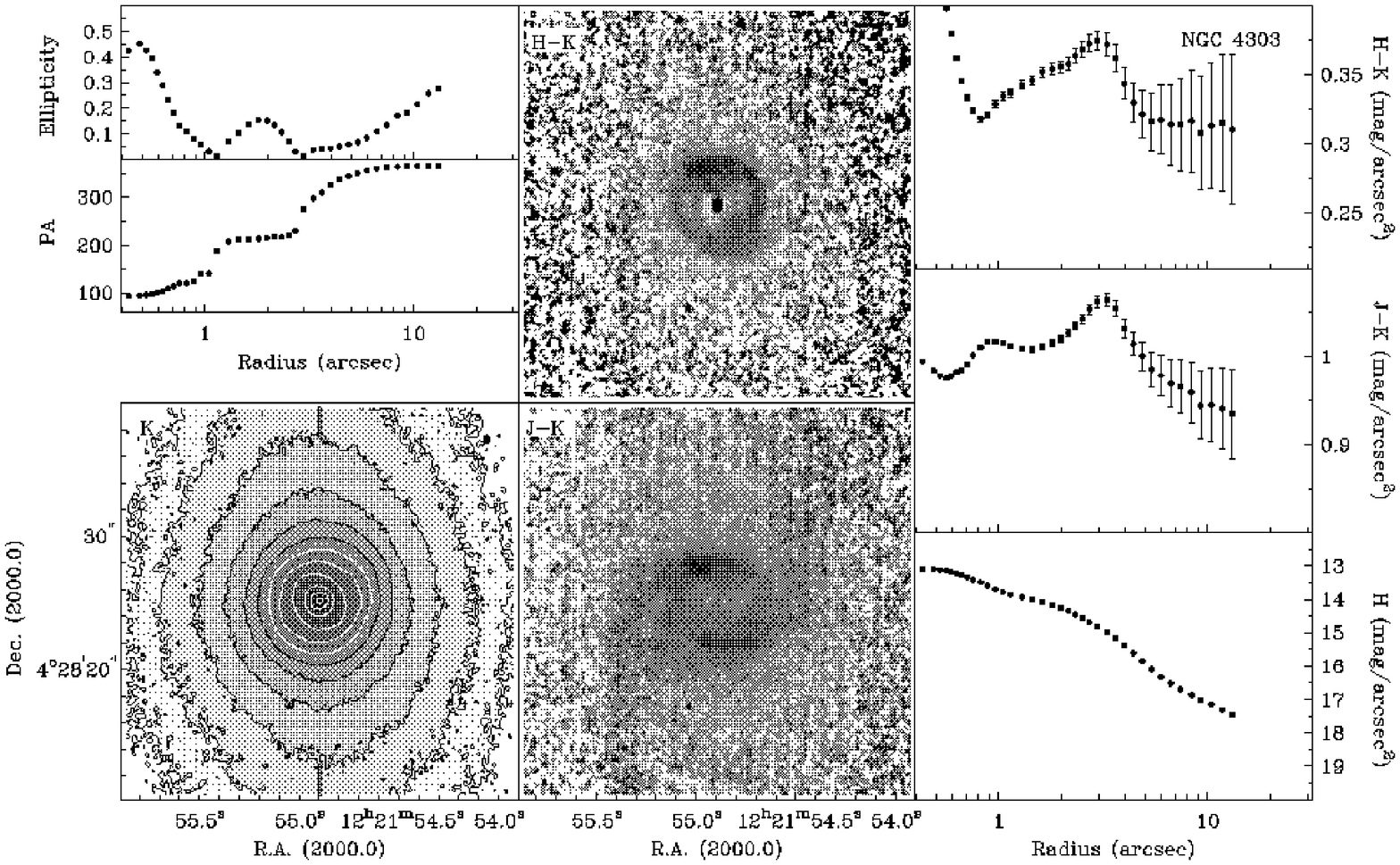,height=12cm,angle=270}
\caption{{\bf h.} NGC 4303}
\end{figure*}

\subsection{NGC 1530}

We find very well-defined mini-spiral structure in the central region of
this galaxy (Fig.~2b).  Reynaud \& Downes (1997; see also Reynaud \&
Downes 1998, 1999) suggest that the molecular gas distribution in the
central 5 kpc of this galaxy is concentrated along arc-like shock
features. Our $J-K$ image shows that these arcs are part of a
well-defined nuclear spiral structure.  Furthermore, Reynaud \& Downes
suggest that a patchy molecular ring may lie inside these shock fronts,
possibly connected to them.  Fig.~2b reveals that the spiral structure
forms a pseudoring around the nucleus.  Even the broad-band image shows
considerable structure in the CNR. No isophote twists can be seen in the
position angle plot. Distinct regions, presumably of enhanced SF, can be
noticed along the mini-spiral in the broad-band NIR images. These
regions coincide in position with the dark (red) lanes seen to outline
the mini-spiral in the $J-K$ image. This implies that the mini-spiral in
$J-K$ is not necessarily a dust spiral (as would have been tempting to
conclude from the colour index image alone) but may well be delineated
by either emission from young stars (e.g. red supergiants), or from hot
dust. We conclude that we see a mini-spiral in the NIR, outlined by
emission from SF and dust. This mini-spiral is well traced also on the
{\it HST} $H$-band image (Fig.~1) as a collection of distinct luminous regions
outlining the armlets, which are accompanied by dust lanes.

Like in the case of NGC~1300, a bump in the radial profiles is visible
at the location of the ring. The $J-K$ colour of the ring is 0.05
magnitudes redder than the background colour. No pronounced change of
slope is detected in the $H$-band surface brightness profile, which
overall decreases more smoothly than in the other galaxies discussed
here.

\setcounter{figure}{1}

\begin{figure*}
\psfig{figure=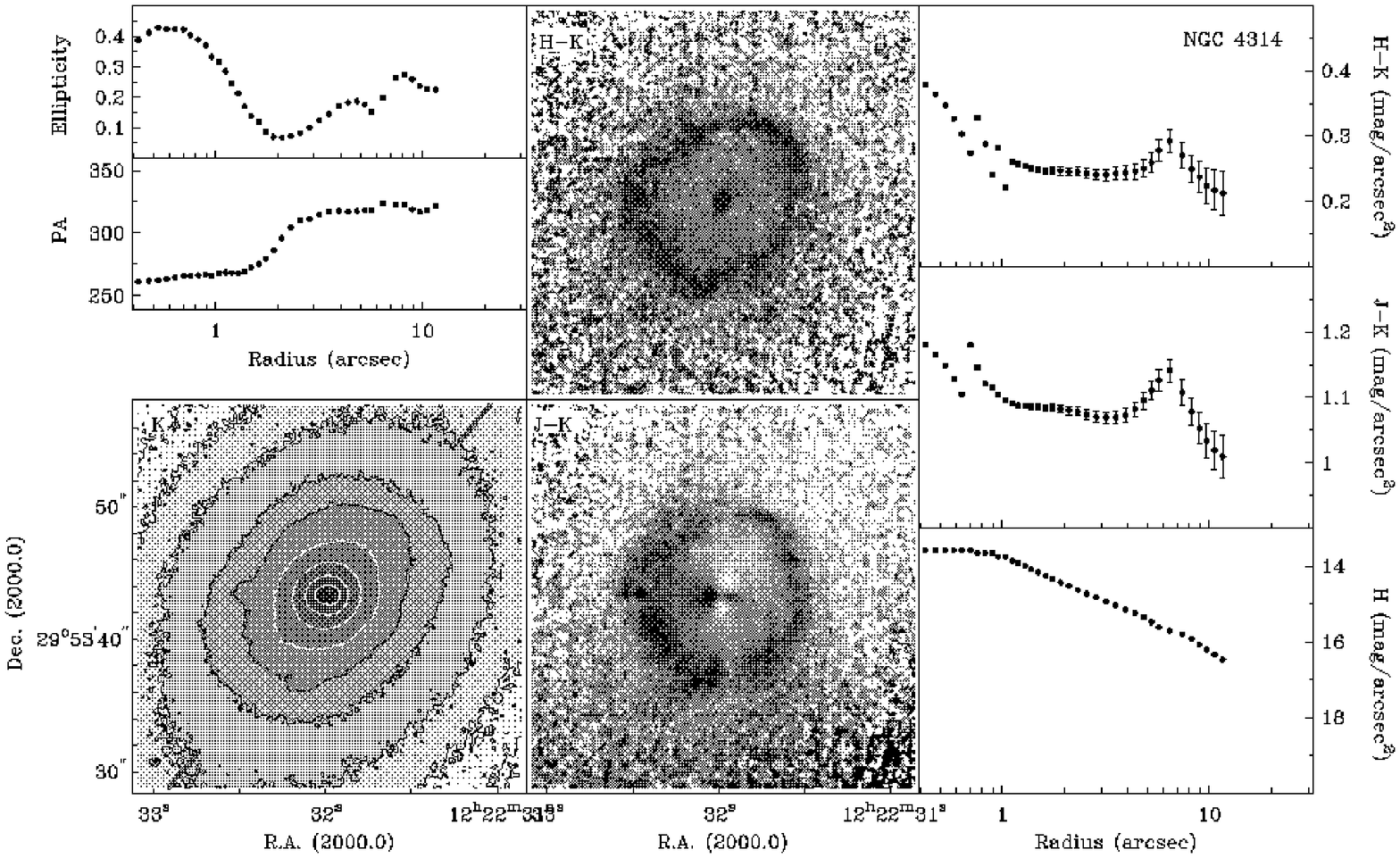,height=12cm,angle=270}
\caption{{\bf i.} NGC 4314}
\end{figure*}

\subsection{NGC 2903}

Fig.~2c shows several peaks of SF in the CNR of this starburst galaxy,
which, as expected, are resolved in much more detail in the {\it HST}
NICMOS image we obtained from the {\it HST} archive (Fig.~1). These
peculiar `hot spots' in the nuclear region have been identified and
described in different ways by various authors. Marcelin, Boulesteix \&
Georgelin (1983) found six `hot spots' forming a linear structure that
crosses the central region.  A later study in infrared and radio by
Wynn--Williams \& Becklin (1985) showed that bursts of SF were not
confined to the visible hot spots. Simons et al. (1988) noticed the
presence of organized dust structure in the CNR using NIR and optical
imaging. We see patches of dust in our images, but no coherent
structure.

Regan \& Elmegreen (1997) found that this galaxy has an isophotal
twist, using a $K$-band image of the inner bar region, although the
results from an earlier study by Elmegreen et al. (1996) were
ambiguous. Similar twists have been interpreted as nuclear bars or
triaxial structures (e.g., Shaw et al. 1995; Friedli \& Martinet 1993;
Wozniak et al. 1995). However, our images show
such an abundance of structure that it is hard to imagine that the
measured changes in the radial behaviour of ellipticity or PA would
have any significance in this respect. For that reason, we set the
ellipticity and the PA to fixed values corresponding to the parameters
of the isophotes on the outskirts of our images when we fit our radial
colour profiles. 

As in, e.g., NGC~1300, the circumnuclear ring-like region of enhanced SF
shows up in the surface brightness profile as a change of slope, and in
the colour profiles as a significant red bump at the same radius.

Our colour index maps suggest that the nuclear hot spots may be part of
a pseudoring. We can identify more than six clumps in the broad-band
images, and even more in the colour index map.  This is in good
agreement with the idea that obscuration by dust is the reason for a
previous underestimate of the extent of this SF region (cf.  Jackson et
al. 1991). As in NGC~1530, we see significant emission from SF regions
in the NIR, including the $K$-band, which must at least in part be due
to red supergiant emission. The red $J-K$ and $H-K$ colours of these
regions makes them stand out clearly in the colour index maps.

\setcounter{figure}{1}

\begin{figure*}
\psfig{figure=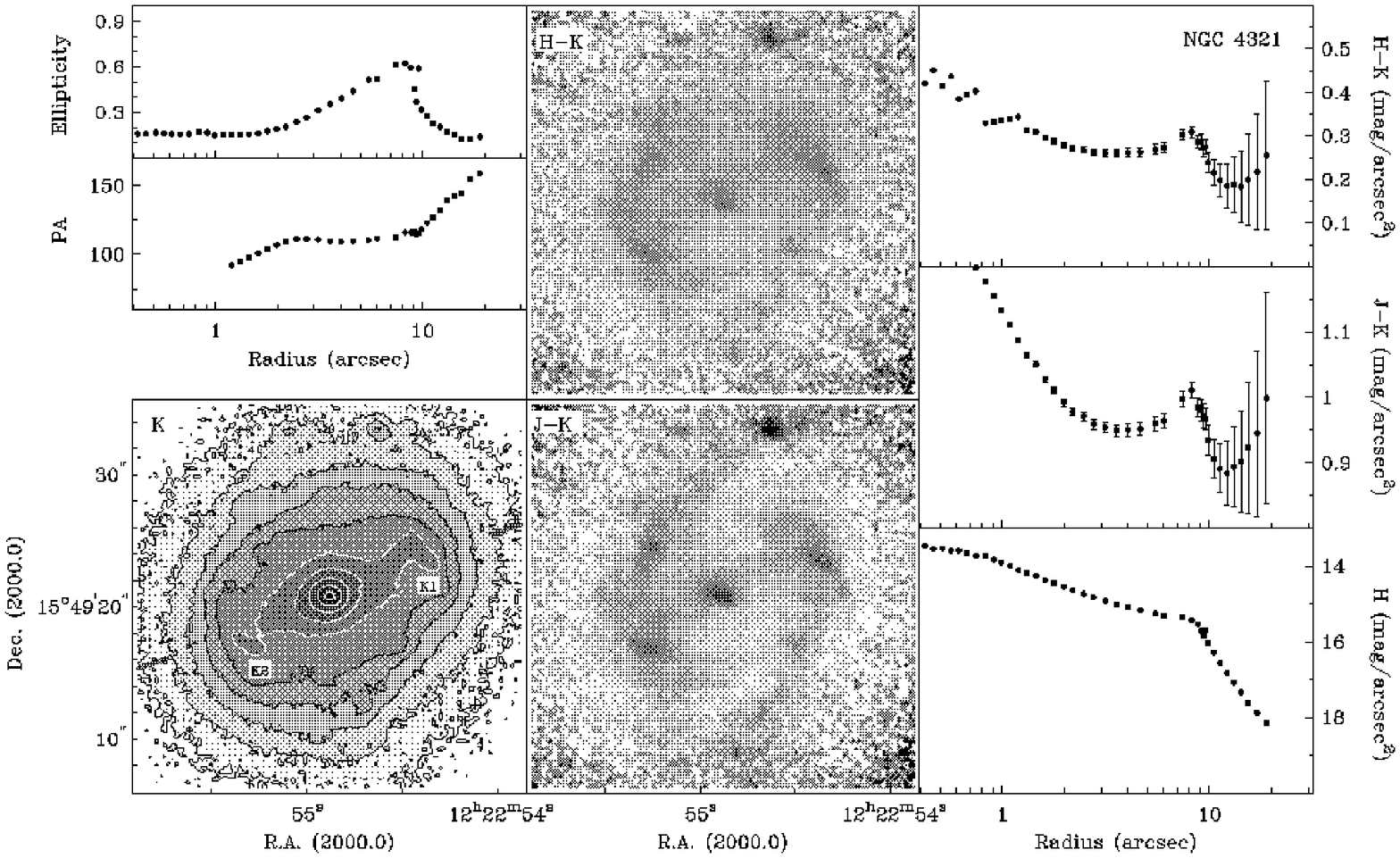,height=12cm,angle=270}
\caption{{\bf j.} NGC 4321}
\end{figure*}

\subsection{NGC 3351}

We only obtained a $K$-band image (Fig.~2d), which shows an incomplete
ring of presumably SF clumps. The {\it HST} $H$-band image (Fig.~1)
shows the same ring but at significantly higher resolution, and resolved
into a large number of individual luminous regions, presumably
star-forming. Some of the brightest knots can be recognised in the UV
($\sim2200$\AA) {\it HST} image published by Colina et al. (1997), but
others, e.g. the one north of the nucleus, are absent from the UV
image. Dust extinction is the most likely candidate mechanism for this
absence, although stellar population differences can be envisaged as
culprits as well.  Spiral arm structure is not obvious in the CNR, and
if it is present at all it is very tightly wound. There is only little
evidence of dust organised into lanes, and the best example of such a
dust lane is seen towards the south-east in the {\it HST} image. In
contrast, and as a direct result of their much lower spatial resolution,
Shaw et al. (1995), found evidence for a circumnuclear ring only in
their $J-K$ colour map.

As in the other galaxies discussed before in this section, there is a
dichotomy in the slope of the $K$-band surface brightness profile.  The
change occurs at the radius of the SF ``ring'', and can be explained as
the transition between an inner active SF region and the quiescent disc
around it. PA and ellipticity are difficult to fit in that area, but
their radial profiles show consistent behaviour in- and outside the
nuclear ring.

\setcounter{figure}{1}

\begin{figure*}
\psfig{figure=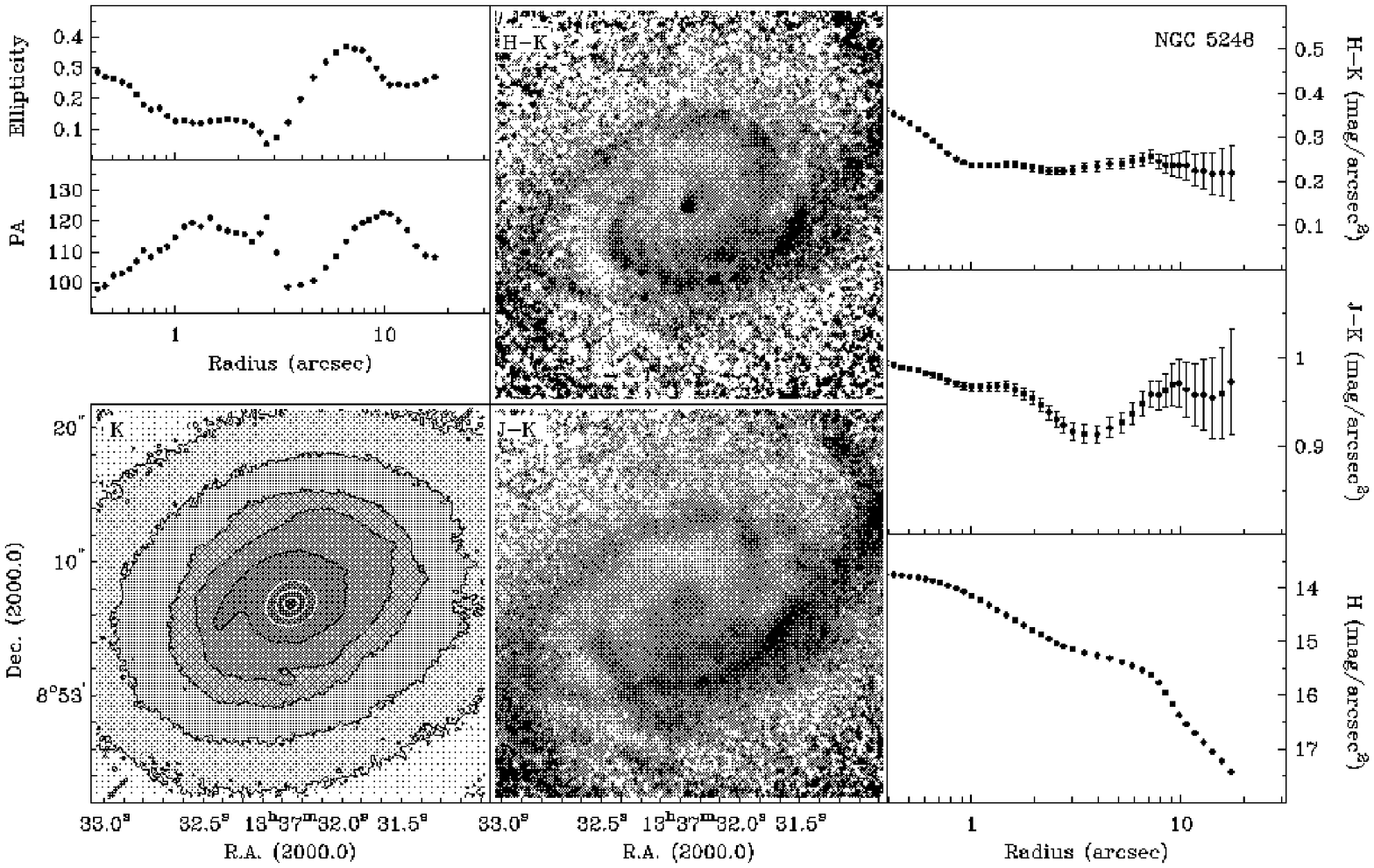,height=12cm,angle=270}
\caption{{\bf k.} NGC 5248}
\end{figure*}\setcounter{figure}{1}

\subsection{NGC 3504}

Our NIR imaging of this barred galaxy reveals a lot of structure,
including a double peak in the central part of the NIR broad-band images
(Fig.~2e).  Recent adaptive optics images obtained with the {\it CFHT} (by
F. Combes \& J.H. Knapen, private communication) confirm this double
peak, but spectroscopic follow-up observations are needed to confirm
whether these peaks are in fact two nuclei. No {\it HST} NIR imaging is
available.

The colour index images show a pair of long and straight dust lanes that
come into the CNR through the main bar, and intricate dust lane
structure in the central region. The nuclear double peak is very
obvious in the colour index maps.  Elmegreen et al. (1997) showed a ring
in $J-K$ with five discrete clumps of SF. The signal to noise ratio in
our images is too low to reveal these SF clumps.

The previously reported isophote twist (e.g., Pompea \& Rieke 1990) is
obvious from our imaging, as well as from the PA and ellipticity
profiles. The double bar, outlined by the two separate peaks in the
ellipticity profile, is most probably not exclusively due to the nuclear
double peak, because the accompanying isophote twist in fact starts
somewhat outside the region of influence of the double peak. The surface
brightness and colour profiles show the hump at the CNR radius that is
seen in many similar galaxies presented in this paper.

\setcounter{figure}{1}

\begin{figure*}
\psfig{figure=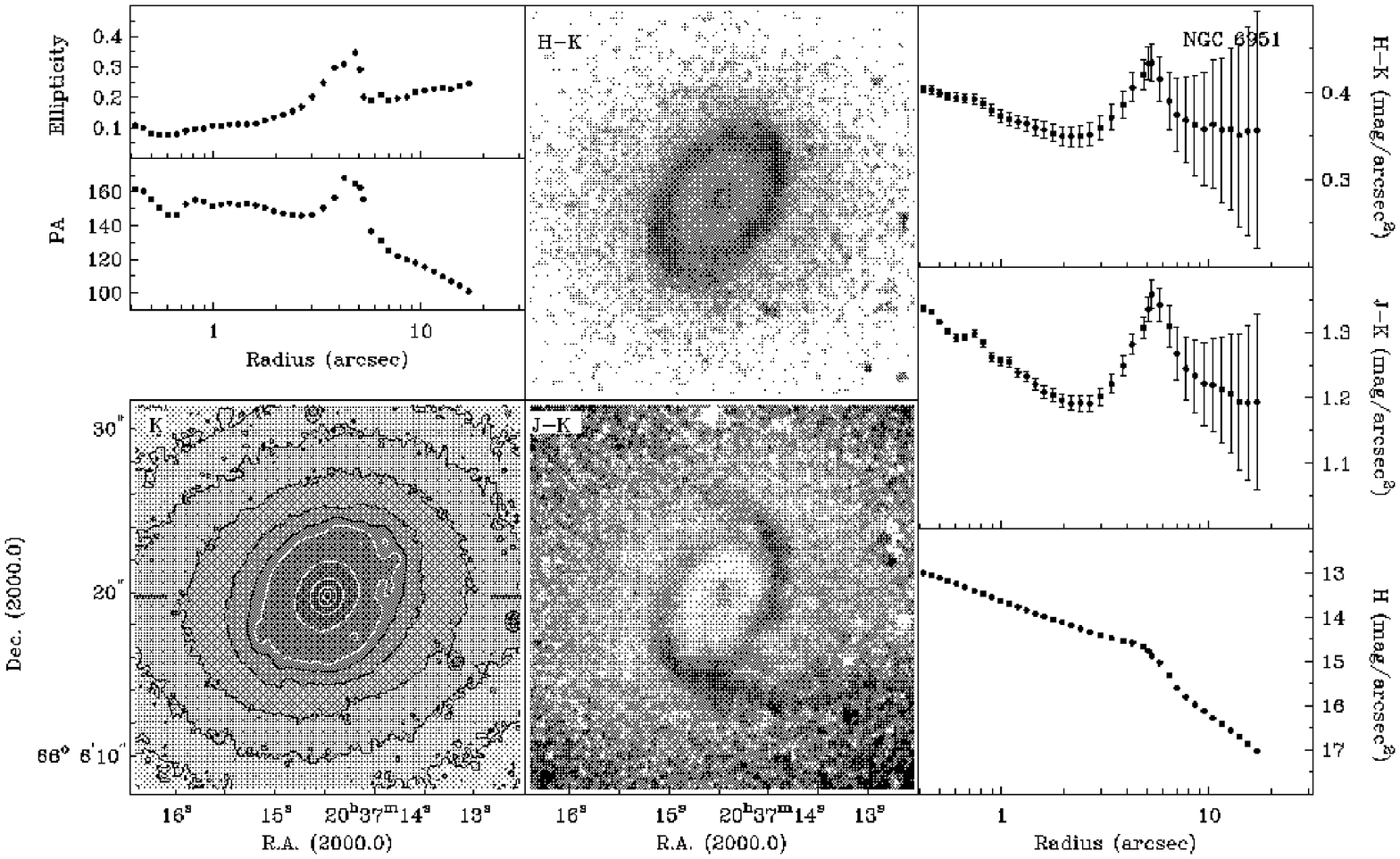,height=12cm,angle=270}
\caption{{\bf l.} NGC 6951}
\end{figure*}

\subsection{NGC 3516}

The only obvious feature in our images is the very red nucleus. Neither
the broad-band {\it HST} $H$-image, nor our ground-based NIR colour maps
reveal any other structure (Fig.~1, 2f). Regan \& Mulchaey (1999) used
an {\it HST} WFPC2--NICMOS colour index map to show that a single spiral
dust pattern dominates the circumnuclear morphology of this galaxy.
They described a strong red dust lane that emerges from a blue feature
north of the nucleus at a radius of 3$\arcsec$. The spatial resolution
of our images is not high enough to reveal such fine details.  The
differences between our images and {\it HST} images are probably due to
the longer spectral baseline that Regan \& Mulchaey used (0.55 $\mu$m to
1.6 $\mu$m) and the different spatial resolution.

Quillen et al. (1999) noticed that the inner $J$-band isophotes are
slightly elongated in a direction roughly perpendicular to the outer
bar, so the galaxy may be doubly barred. Our $H$-band profile indeed
shows an isophote twist of about 50\degr\ (Fig.~2f). Quillen et
al. (1999) saw a curved dust feature at about 4 arcsec south of the
nucleus in their {\it HST} WFPC2--NICMOS colour index map.

\subsection{NGC 3982}

We see a multi-armed spiral pattern in our images (Fig.~2g), in
agreement with Regan \& Mulchaey (1999) who describe the global
morphology of this galaxy as multi-armed. There is a lot of structure in
our ground-based images, but a detailed comparison with the colour index
maps obtained by Regan \& Mulchaey (1999) is difficult due to the lower
spatial resolution of our images. The {\it HST} NIR image (Fig.~1) is the one
used by Regan \& Mulchaey, and shows the rather faint spiral structure
in the CNR. Our colour index map shows that this galaxy has a small red
nucleus, classified as Sy2.  The ellipticity of the isophotes reaches a
maximum at a radius of 9 arcsec.  This could correspond to a ring, a
small bar or a triaxial bulge.

The radial colour profiles ($J-K$ and $H-K$) follow a characteristic
shape, becoming very red close to the nucleus, with the colour most
likely due to emission from dust heated by the AGN radiation field. The
colours become bluer until a certain radius (the location of the ring),
after which they remain constant. Such a profile shape is only seen in
the AGNs of our sample. The difference in colour between the nucleus and
the ring radius is about 0.2 magnitudes in both $J-K$ and $H-K$.

\subsection{NGC 4303}

Buta \& Crocker (1993) classify this galaxy as having a nuclear ring,
based on their H$\alpha$ data. Elmegreen et al. (1997), using NIR
observations, did not detect any ring. According to them, the ring
consists of very young stars which do not show up in the NIR. However,
we can see a well-defined ring in our $J-K$ and $H-K$ images
(Fig.~2h). There is a pair of dust lanes which connect the bar to the
nuclear ring in the south and north. The reddest colours are seen
where the dust lanes merge with the nuclear ring. 

Colina et al. (1997) present a UV ($\sim2200$\AA) {\it HST} image of the
CNR of NGC~4303, which shows spiral structure outlining massive SF,
continuing all the way into the unresolved core on the NE side of the
nucleus. The UV SF spiral is strongest on the side opposite to where we
see the largest concentrations of dust (darkest patches in Fig.~2h) in
our NIR colour index maps. Our {\it HST} $H$-band image shows some
spiral structure in the NIR but emission is dominated by the central
bulge component.

There are strong isophotal twists within the central 7 arcsec, and a
possible nuclear bar with a radius of 2 arcsec where the ellipticity
reaches a peak of 0.2 and the position angle is constant at about
220\degr. The signature of the ring can be recognised as a red peak in
the $J-K$ and $H-K$ colour profiles, as well as in all other profiles.

\subsection{NGC 4314}

Our ellipticity and PA profiles provide some evidence for a nuclear bar
with a radius of 1--2 arcsec. The corresponding PA twist can be
recognized in the contours of the broad-band image.  Benedict et
al. (1993), using the WFPC camera on the {\it HST}, found an oval
distortion with a length of 8 arcsec in the nuclear region. Wozniak et
al. (1995) suggested that this galaxy could be another example of a
double-barred galaxy, although they found it difficult to confirm this
with their ground-based image at a considerably lower spatial resolution
than our new data.

The colour maps published by Shaw et al. (1995) revealed the presence of
the circumnuclear ring, which is so prominent in the optical (e.g.,
Morgan 1958), and especially in H$\alpha$ (e.g., Pogge 1989a). This ring
is redder than all the other regions in the galaxy, having colours
consistent with those of typical old stellar populations in ellipticals
and spiral bulges. We see a smooth and continuous ring in our $J-K$
colour index map (Fig.~2i). The {\it HST} $H$-image (Fig.~1) shows a wealth of
emitting structure in narrow tightly wound spiral armlets in the nuclear
ring. Fine dust lanes are seen to accompany the spirals.  The radial
profiles also show the signature of the ring, most clearly with colours
that are about 0.1 magnitude redder than the background.

\subsection{NGC 4321}

The prominent star-forming CNR of NGC 4321 (M100) has been studied in
great detail using, e.g., optical and NIR imaging, CO interferometry,
and modelling (e.g., Pogge 1989; Knapen et al. 1995a,b; Knapen 1998;
Wada et al. 1998; Garc\'\i a--Burillo et al. 1998; Ryder \& Knapen 1999;
Knapen et al. 2000).  Knapen et al. (1995b) concluded that M100 has a
circumnuclear starburst maintained by a global bar-driven density wave.

As already shown in detail by Knapen et al. (1995a), the $K$-image of
this region is generally smooth, in contrast to the appearance in
optical and H$\alpha$. Two symmetrically placed `hot spots', named K1
and K2, are obvious in Fig.~2j, both in emission in the broad-band
image, and as red features in the colour index images. Ryder \& Knapen
(1999) recently used NIR imaging and spectroscopy to confirm the
suspicion that K1 and K2 are in fact regions of enhanced SF, and the $K$
emission from those regions is partly due to young stars (Knapen et
al. 1995a,b). Our NIR imaging confirms the location of dust lanes and
suspected SF regions, shown by Knapen et al. (1995a) in their $I-K$
colour index map. The locus of the circumnuclear ring-like structure
shows up prominently in all radial profiles. Unfortunately, no {\it HST} NIR
images are available.

\subsection{NGC 5248}

NGC 5248 is a galaxy with an H{\sc ii} nucleus and a lot of SF activity
in the CNR. Elmegreen et al. (1997) found very conspicuous central
spiral arms, and several hotspots that form a ring-like spiral pattern.
Buta and Crocker (1993) detected a nuclear ring with a diameter of
10--17 arcsec.  This activity shows up clearly in our broad-band NIR
images, as well as in our colour index maps (Fig.~2k). Spiral structure
with star-forming arms, accompanied by dust lanes, is the dominant
feature. The western spiral arm has colours which are redder by about
0.1 mag in $J-K$ than its counterpart in the east. Our $J-K$ image has
been published earlier by Laine et al. (1999). They compared it with the
images obtained using adaptive optics, which show a nuclear grand-design
spiral structure. This nuclear spiral, at scales of tens of pc, is not
expected to show up in a single broad-band NIR image, even at {\it HST}
resolution, and in fact does not show up (Fig.~1). The {\it HST} $H$-band
image does show a wealth of structure in the CNR, again in the form of
emitting regions distributed along spiral arm fragments, and accompanied 
by less luminous regions which may well be dusty.

As in other galaxies, we can see the signature of the ring as a peak at
a radius of $\sim7$ arcsec in all radial profiles. There is no evidence
for nested bars.

\subsection{NGC 6951}

The dust structure in the circumnuclear ring is the most conspicuous
feature in our $J-K$ colour index map, but the star-forming regions in
the ring can also be seen in the broad-band images
(Fig.~2l). The bar dust lanes connect to this nuclear ring in the
northeast and southwest. Several sites of SF are located along the
ring, and its presence is also seen in the ellipticity and PA
profiles, as well as in all other profiles (Fig.~2l).

We can also see a blue ring at about 2 arcsec, within the red nuclear
ring surrounding it. This blue ring shows up as a dip in the $J-K$
profile. Its nature is not clear, and needs further study. The {\it HST}
$H$-band image, used also by P\'erez et al. (2000), shows a picture also
seen in, e.g., NGC~3351 and NGC~4314, namely of bright emitting knots
distributed along tightly wound spiral armlets in the nuclear ring,
accompanied by dust lanes.

\section{Discussion}

We have presented examples of circumnuclear rings and dust lane patterns
in the CNRs of a dozen barred galaxies. In most galaxies, rings are well
defined, in others, nuclear spirals with different curvature radii are a
predominant circumnuclear feature. There are also some galaxies where
both features coexist.  Our sample is composed of 12 galaxies, most of
which have nuclear rings. Six of them also have clear nuclear
mini-spirals (NGC~1530, NGC~3504, NGC~4314, NGC~4321, NGC~5248 and
NGC~6951), while in a few others there is less clear or circumstantial
evidence for mini-spirals.

\subsection{Circumnuclear Rings}

Simulations (e.g. Athanassoula 1992; Byrd et al. 1994; Heller \&
Shlosman 1994; Knapen et al. 1995b; Piner, Stone \& Teuben 1995; see
also review by Shlosman 1999) have shown that circumnuclear rings can
arise as a consequence of bar-driven inflow and the existence of
dynamical resonances in the bar. Shocks form in the gas in regions of
orbit crowding along the leading edges of the bar, and dense gas
accumulates in these regions. In the shock region the gas loses angular
momentum through torques exerted by the bar, and flows inward. The
pattern of the gas flow can be quite complex, but if two ILRs are
present, inflowing gas accumulates in a ring between them (reviewed by
Shlosman 1999). If there is no ILR, the gas may continue to flow inward,
resulting in a nuclear starburst rather than a ring (Telesco, Dressel \&
Wolstencroft 1993). Other mechanisms for the origin of circumnuclear
rings that have been put forward as possible origins include shear in
the differentially rotating disc (Buta \& Combes 1996), supermassive
black hole binaries (Taniguchi \& Wada 1996), and minor mergers
(Taniguchi 1999).

Our sample galaxies were selected to have some kind of circumnuclear
structure, and except in NGC~3516 and NGC~3982, all have clear
circumnuclear star-forming ring-like regions. SF is visible in most
galaxies in the high-resolution {\it HST} NICMOS images (Fig.~1), and
also in a number of galaxies in the ground-based images. However, in all
galaxies with circumnuclear SF, dust lane structure outlining
small-scale spiral arms is present, as can be seen in the colour index
maps (Fig.~2). This is true even for NGC~1300 where the SF regions are
conspicuously absent from our ground-based NIR data and very weak in
{\it HST} NIR images. We conclude that in this class of barred galaxies
with circumnuclear SF, dust lane structure in the CNR is always direct
evidence of the accompanying SF, which may also show up in broad-band
NIR imaging.

NIR data are not optimal for determining the sizes of the
circumnuclear features, and we will come back to this issue in future
papers, where \ha\ imaging will be used to localize the SF activity in
the CNR. However, the radial profiles often allow the determination of
the diameter of the ``rings'': typical major axis diameters are 1-2
kpc.

\subsection{Dust Lanes}

The first suggestion about the link between dust lanes and shocks in the
gas flow was made by Prendergast (1962). Gas can follow simple periodic
orbits that do not intersect when the bar or oval distortion is not
very strong. When the bar is strong, the families of periodic orbits
intersect and shocks are formed (e.g. Sanders, Teuben and Van Albada
1983).

Dust lanes in bars, as described by Athanassoula (1992),
can be:

\begin{enumerate}
\item straight and parallel to the bar major axis, or
\item curved and have their concave sides towards the bar major
axis.
\end{enumerate}

The detailed shape of the loci of offset shocks changes with the
parameters of the potential (Athanassoula 1992). The simulations
presented by Athanassoula show that there is a very clear sequence of
shapes for the shock loci. For axial ratios corresponding to fat bars or
ovals, and presumably to what could theoretically be called a weak bar,
the shock lanes are curved with their concave parts towards the bar
major axis. For slim (``strong'')bars, the dust lanes straighten out,
their curvature decreases and their shape becomes gradually more
straight. For even higher axial ratios the shock loci tend towards the
bar major axis. Thus, strong bars should have straight dust lanes, while
ovals or weak bars should have more curved dust lanes.

Athanassoula (1992) also proposed that the strength of SF near the dust
lanes depends on the strength of the bar. The dust lanes observed on the
leading edges of most bars with high ellipticity (strong) are very
smooth and show no H{\sc ii} regions or other signs of recent SF. In the
case of curved dust lanes strings of H{\sc ii} regions are accompanying
the dust lanes. Due to the enhanced gas mass fraction, dust is expected
to be prominent in the CNRs themselves, and this, in fact, is seen
clearly in most cases in the colour index images, and also often in the
broad-band images, especially the {\it HST} NICMOS images. Such dust
lanes are expected to be continuations of the dust lanes in the bar, and
this is what is seen in our images (see Shlosman 1999 for a
theoretical review).

NIR colour index images are not very sensitive to changes in stellar
populations, and can outline dust structure clearly (see e.g. the $I-K$
map of M100 in Knapen et al. 1995a). We see clear and abundant
observational evidence for dust lanes on several scales, but most
clearly in the CNRs. Dust lanes in the bars are not well visible in
general in our NIR imaging due to the lower signal to noise ratios
achieved in the bar regions.  In Paper II, we present optical colour
index maps which outline the dust lane structure in the bars and discs
of our sample galaxies more clearly, and we study the relationship
between the shape of the dust lanes, the axial ratio, and the SF in the
bar in more detail.

\subsection{Colour Distributions}

Our colour maps have shown the presence of circumnuclear rings in most
of the galaxies in our sample. These rings are generally redder than
other regions in the galaxy by about 0.05-- 0.1 magnitude in $J-K$ and
$H-K$. From the colour index maps or colour profiles alone it is not
possible to make meaningful statements about quantities of
extinguishing dust implied by the redder colours. In fact, there are
indications from both optical and NIR imaging and spectroscopy that the
red colours in CNRs in galaxies like the ones studied here may be influenced  
by young stars, e.g. red supergiants (Knapen et al. 1995a,b;
Knapen 1996; Elmegreen et al. 1997; Ryder \& Knapen 1999). In Paper
III, we will compare the precise location of the $J-K$ and $H-K$
features with those of SF regions as seen in \ha\ emission, and try to
place quantitative limits on the origins of the red light in the CNRs.

Two of the host galaxies of Sy nuclei (NGC~3516 and NGC~3982), which
have colours consistent with those measured by Peletier et al. (1999),
also show a peculiar shape in the colour profiles, starting from a very
red value and steeply decreasing until the radius of the ring, remaining
constant afterwards.  Peletier et al. (1999) suggested that the red
$H-K$ (or $J-K$) colours in the cores of many Sy galaxies could be due
to a significant fraction of thermal radiation from hot dust heated by
the Sy nucleus.  However, the two other AGN hosts in our sample
(NGC~4303 and NGC~6951) do not show significantly red nuclei.

\subsection{Radial profiles}

One of the main data products presented in this paper is a set of
detailed radial profile fits to the NIR images. We present (Fig.~2a-l)
radial profiles of ellipticity, major axis position angle, NIR colour,
and $H$-band surface brightness. We will discuss the existence and
properties of (primary and secondary) bars in Paper II, where we combine
the NIR profiles presented here with $I$-band profiles of the complete
discs, covering also the primary bar, if present.  Certain aspects of
the radial profiles have already been discussed in Section~3, but there
is one characteristic which is worth mentioning here. Apart from
NGC~3516 and NGC~3982, the radial profiles show characteristic behaviour
near the radius of the circumnuclear ``ring''. In seven (out of 10)
galaxies, {\it all} profiles (surface brightness, colour, ellipticity
and position angle) show either a change in slope, where the surface
brightness profile becomes steeper, or a bump, when the colour profile
shows a limited redder region caused by either dust or SF in or near the
ring -- see Section 4.3. These characteristics are accompanied by some
change in ellipticity and position angle. In a further two galaxies
(NGC~1530 and NGC~4314) only the surface brightness profile does not
follow this trend, whereas in NGC~5248 the colour profile is the deviant
one. Of course in NGC~2903 and NGC~3351 we cannot judge the behaviour of
ellipticity/position angle and colour profiles, respectively.

We postulate that the features in the surface brightness and colour
profiles are due directly to a relatively young population. Firstly, an
old, bulge-like, population would not be expected to give rise to strong
high-resolution features such as the ones discussed here.  Secondly, if
the red bumps in the colour profiles were due to extinction by dust
only, they should be accompanied by relative dips in the corresponding
surface brightness profiles (such dust would remove light in each of the
$J, H$ and $K$-bands). Instead, red bumps in the colour profiles are in
all cases accompanied by bumps in the surface brightness profiles:
relative excesses in emission, red in colour, that are naturally
explained by excess emission produced locally by relatively young stars,
probably accompanied by dust.  We will come back to this issue in Paper
III, where we combine the NIR data with \ha\ imaging.

\section{Conclusions}

This is the first paper in a series exploring the relationships between
ring-like star-forming regions in the central few kpc regions of barred
galaxies and the morphological properties of their hosts. In this paper,
we present new sub-arcsec resolution NIR images obtained with the {\it
CFHT} of the central regions of a sample of 12 barred galaxies with
circumnuclear SF activity, and {\it HST} archival NIR images of most of
the sample objects. In Paper~III, we will combine these data with
optical imaging of the complete hosts, as presented in Paper II.

Our NIR images reveal a wealth of structure, caused by young stars and
dust.  In colour index maps, structures such as SF sites and dust lanes
are generally apparent, even in those galaxies where the broad-band
images appear featureless. The colour maps of most galaxies show
evidence of an often tightly wound spiral structure. In some cases, this
spiral structure is visible also in the broad-band images, and as
expected due to the higher spatial resolution, more often so in the {\it
HST} than in the ground-based images. Where circumnuclear structure is
visible in the {\it HST} images (in about three quarters of all cases),
it shows large numbers of small emitting regions, presumably SF regions,
often outlined in spiral arm fragments, and accompanied by dust
lanes. Circumnuclear spiral structure thus appears to be common in
barred spiral galaxies with circumnuclear SF.

In most of our sample galaxies, and in all those where the star-forming
nuclear ring is well defined, radial profiles of surface brightness,
colour, ellipticity and position angle show a characteristic bump or
change in slope at the radius where the circumnuclear ring is located. 

{\it Acknowledgements} This research has made use of the NASA/IPAC
Extragalactic Database (NED) which is operated by the Jet Propulsion
Laboratory, California Institute of Technology, under contract with the
National Aeronautics and Space Administration.

\bsp
\label{lastpage}


\begin{thebibliography}{99}

\bibitem{} Aaronson M., 1977, PhD thesis, Harvard Univ.

\bibitem{} Athanassoula E., 1992, MNRAS, 259, 345

\bibitem{} Balzano V. A., Weedman D. W., 1981, ApJ, 243, 756

\bibitem{} Benedict G. F., Higdon J. L. et al., 1993, AJ, 105, 1369

\bibitem{} Buta R., Combes F., 1996, Fund. Cosmic Phys., 17, 95

\bibitem{} Buta R., Crocker D. A., 1993, AJ, 105, 1344

\bibitem{} Byrd G., Rautiainen P., Salo H., Buta R., Crocher D. A., 
1994, AJ, 108,476

\bibitem{} Cizdziel P. J., Wynn--Williams C. G., Becklin E. E., 
1985, AJ, 90, 731

\bibitem{} Colina L., Garc\'\i a--Vargas M. L., Mas--Hesse J. M., 
Alberdi A., Krabbe A., 1997, ApJ, 484, L41

\bibitem{} Devereux N. A., Becklin E. E., Scoville N., 1987, ApJ, 312, 529

\bibitem{} de Vaucouleurs  G., de Vaucouleurs A., Corwin H. G. Jr., 
Buta R. J., Paturel G., Furgue P., 1991, Third Reference Catalogue of
Bright Galaxies.  Springer, New York (RC3)

\bibitem{} Elmegreen, B.G., 1994, ApJ, 425, L73

\bibitem{} Elmegreen, B.G., 1997, Rev. Mex. Astr., 6, 165

\bibitem{} Elmegreen D. M., Chromey F. R., Santos M., Marshall D., 1997, 
AJ, 114, 1850

\bibitem{} Elmegreen D. M., Elmegreen B. G., Chromey F. R., Hasselbacher 
D. A., Bissell, B.A., 1996, AJ, 111, 1880

\bibitem{} Ferrarese, L. et al. 1996, ApJ 464, 568

\bibitem{} Friedli D., Martinet L., 1993, A\&A, 277, 27

\bibitem{} Garc\'\i a--Burillo S., Sempere M. J., Combes F., Neri R., 
1998, A\&A 333, 864

\bibitem{} Glass I. S., 1976, MNRAS, 175, 191
 
\bibitem{} Glass I. S., 1984, MNRAS, 211, 461

\bibitem{} Graham J.A. et al., 1997, ApJ 477, 535

\bibitem{} Heller C. H., Shlosman I., 1994, ApJ, 424, 84

\bibitem{} Hunt L. K., Zhekov S., Salvati M., Mannucci F., Stanga R. M.,
 1994, A\&A, 292, 67

\bibitem{} Jackson J. M., Eckart A., Cameron M., Wild W., Ho P. T. P., 
Pogge R. W., Harris A. I., 1991, ApJ, 375, 105

\bibitem{} J\o rgensen I., Franx M., Kjaergaard P., 1992, A\&AS, 95, 489

\bibitem{} Knapen J. H., 1996, in Proc. Nobel Symposium 98
{\it Barred Galaxies and Circumnuclear Activity}, Eds. Aa. Sandqvist,
P.O. Lindblad, Lecture Notes in Physics, 474, 233, Springer-Verlag,
Berlin

\bibitem{} Knapen J. H., 1998, MNRAS, 297, 255

\bibitem{} Knapen J. H., 1999, in Beckman J. E., Mahoney T. J., eds, The 
Evolution of Galaxies on Cosmological
Timescales. Astron. Soc. Pac., San Francisco, Vol. 187, 72

\bibitem{} Knapen J. H., Beckman J. E., Shlosman I., Peletier R. F.,  
Heller C. H., de Jong R. S., 1995a, ApJL, 443, L73

\bibitem{} Knapen J. H., Beckman J. E., Heller C. H., Shlosman I.,
de Jong R.S., 1995b, ApJ, 454, 623

\bibitem{} Knapen J. H., Shlosman I., Heller C. H., Rand R. J.,
Beckman J. E. \& Rozas, M., 2000, ApJ, 528, 219

\bibitem{} Laine S., Knapen J. H., P\'erez--Ram\'\i rez D., Shlosman 
I., Doyon R., Nadeau D., 1999, MNRAS, 302, L33.

\bibitem{} McAlary C. W., McLaren R. A., Crabtree D. R., 1979, ApJ 234, 471

\bibitem{} Maoz D., Barth A. J., Sternberg A., Filippenko A. V., Ho L. C.,
Macchetto F. D., Rix H.--W., Schneider D. P., 1996, AJ, 111 2248

\bibitem{} Marcelin M., Boulesteix J., Georgelin Y., 1983, A\&A, 128, 140

\bibitem{} Martin P., Friedli D., 1997, A\&A, 326, 449

\bibitem{} Morgan, W.W., 1958, PASP, 70, 364

\bibitem{} Nadeau D., Murphy D. C., Doyon R., Rowlands N., 1994, PASP, 106, 909

\bibitem{} Peletier R. F., Knapen J. H., Shlosman I., P\'erez--Ram\'\i rez D.,
Nadeau D., Doyon R., Rodriguez Espinosa J. M., P\'erez Garc\'\i a
A. M., 1999, ApJS, 125, 363 

\bibitem{} P\'erez, E., M\'arquez, I., Marrero, I., Durret, F.,
Gonz\'alez Delgado, R.M., Masegosa, J., Maza, J., Moles, M., 2000, A\&A
353, 893

\bibitem{} P\'erez--Ram\'\i rez D., Knapen, J.H., 2000a, MNRAS,
submitted (Paper~II)

\bibitem{} P\'erez--Ram\'\i rez D., Knapen, J.H., 2000b, MNRAS, in
preparation (Paper~III)

\bibitem{} Phinney E. S., 1994, in Shlosman I., ed, Mass-Transfer Induced 
Activity in Galaxies. Cambridge: Cambridge Univ. Press, 1

\bibitem{} Piner B. G., Stone J. M., Teuben P. J., 1995, ApJ, 449, 508

\bibitem{} Pogge R. W., 1989a, ApJS, 71, 433

\bibitem{} Pogge R. W., 1989b, ApJ, 345, 730

\bibitem{} Pompea, S.M., Rieke, G.H., 1990, ApJ, 356, 416

\bibitem{} Prendergast, K.H., 1962, in Distribution and Motion of ISM in 
Galaxies, ed. L. Wotljer (New York: Benjamin), 217

\bibitem{} Quillen A. C., Alonso-Herrero A., Rieke M. J., McDonald C.,
Falcke H., Rieke, G.H. 1999, ApJ, 525, 685

\bibitem{} Regan M. W., Elmegreen D. M., 1997, AJ, 114, 965

\bibitem{} Regan M. W., Mulchaey J. S., 1999, AJ, 117, 2676

\bibitem{} Reynaud D., Downes D., 1997, A\&A, 319, 737

\bibitem{} Reynaud D., Downes D., 1998, A\&A, 337, 671

\bibitem{} Reynaud D., Downes D., 1999, A\&A, 347, 37

\bibitem{} Ryder S. D., Knapen J. H., 1999, MNRAS, 302, L7

\bibitem{} Sanders R. H., Teuben P. J., van Albada G. D., 1983, 
in Athanassoula E., ed, Proc. IAU Symp. 100, Internal kinematics and 
dynamics of galaxies. Reider, Dordrecht, p. 211

\bibitem{} Sersic J. L., Pastoriza M. G., 1967, PASP, 79, 152

\bibitem{} Shaw M. A., Axon D. J., Probst R., Gatley I., 1995, MNRAS, 274, 369

\bibitem{} Shlosman I., 1999, in Beckman J. E., Mahoney T. J., eds, 
The Evolution of Galaxies on Cosmological Timescales. Astron. Soc. Pac., 
San Francisco, Vol. 187, 100

\bibitem{} Shlosman I., Frank J., Begelman M. C., 1990, Nature, 345, 679

\bibitem{} Simons D. A., Depoy D. L., Becklin E. E., Capps R. W.,
Hodapp K.--W., Hall D.N.B., 1988, ApJ, 335, 126

\bibitem{} Spinoglio L., Malkan M. A., Rush B., Carrasco L., Recillas--Cruz 
E., 1995, ApJ, 453, 616

\bibitem{} Taniguchi Y., 1999, ApJ, 524, 65

\bibitem{} Taniguchi Y., Wada K., 1996, ApJ, 469, 581

\bibitem{} Telesco C. M., Dressel L. L., Wolstencroft R. D., 1993, ApJ, 414,120

\bibitem{} Wada K., Sakamoto K., Minezaki T., 1998, ApJ, 494, 236

\bibitem{} Willner S. P., Elvis M., Fabbiano G., Lawrence A., Ward M. J., 
1985, ApJ, 299, 443

\bibitem{} Wozniak H., Friedli D., Martinet L., Martin P., Bratschi
P., 1995, A\&AS, 111, 115

\bibitem{} Wynn--Williams C. G., Becklin E. E., 1985, ApJ, 290, 108

\end{thebibliography}
\end{document}